# Changing spatial distribution of water flow charts major change in Mars' greenhouse effect

Edwin S. Kite[1,*], Michael A. Mischna[2], Bowen Fan[1], Alexander M. Morgan[3,4], Sharon A. Wilson[3], Mark I. Richardson[5].

[1] University of Chicago, Chicago, IL, 60637.
[2] Jet Propulsion Laboratory, California Institute of Technology, Pasadena, CA, 91109.
[3] Smithsonian Institution, Washington, DC, 20002.
[4] Planetary Science Institute, Tucson, AZ, 85719.
[5] Aeolis Research, Chandler, AZ, 85224.

* Correspondence to: kite@uchicago.edu

**Abstract**
Early Mars had rivers, but the cause of Mars' wet-to-dry transition remains unknown. Past climate on Mars can be probed using the spatial distribution of climate-sensitive landforms. We analyzed global databases of water-worked landforms and identified changes in the spatial distribution of rivers over time. These changes are simply explained by comparison to a simplified meltwater model driven by an ensemble of global climate model simulations, as the result of ≳10 K global cooling, from global average surface temperature $\bar{T} \geq 268$ K to $\bar{T} \sim 258$ K, due to a weaker greenhouse effect. In other words, river-forming climates on Early Mars were warm and wet first, and cold and wet later. Surprisingly, analysis of the greenhouse effect within our ensemble of global climate model simulations suggests that this shift was primarily driven by waning non-$CO_2$ radiative forcing, and not changes in $CO_2$ radiative forcing.

**Introduction**
Mars' climate 3.6-3.0 Ga was at least occasionally warm enough for rivers and lakes (inferred to have been habitable) (e.g. *1-4*), but the surface today is a cold desert. Few constraints exist on Mars' atmospheric greenhouse effect during the wet-to-dry transition (*4*). Was Early Mars temperate or icy, was the environmental catastrophe abrupt or gradual, and what caused the change? The prevailing view is that the cause of Mars' drying-out was loss of atmospheric $CO_2$. Indeed, Mars' atmosphere today is so thin (~6 mbar, $CO_2$-dominated) that it is close to the triple point of water, so lakes on early Mars probably formed under a thicker atmosphere (*5-7*). $CO_2$ is, in the modern inner Solar System, a key greenhouse gas for regulating climate change (*8*). However, even when $H_2O$-vapor feedback is considered, additional non-$CO_2$ warming is needed to warm early Mars enough for rivers (*9*). Therefore, changes in non-$CO_2$ radiative forcing are an alternative explanation for Mars' wet-to-dry transition. The relative importance of these two mechanisms has not been investigated, so the prevailing explanation of the wet-to-dry transition remains untested.

Here, we reconstruct the history of Mars' greenhouse effect using geologic proxies for past river activity that time-resolve Mars' desertification. We compare the proxy data to a climate model in order to retrieve changes in the greenhouse effect, and also to assess the extent to which the changes were the result of changes in $CO_2$ radiative forcing versus non-$CO_2$ radiative forcing. Finally, we discuss the implications for the cause of the cooling and drying-out of Mars.

Our approach uses Mars' geologic record of precipitation-fed water runoff (meltwater and/or rain) spanning multiple eras (e.g., *3-4*, and references therein) (Fig. S2). Early on (>3.6 Ga), spatially pervasive and regionally integrated Valley Networks formed (*10*). Later, 3.6-3.0 Ga (and perhaps still later), spatially patchy alluvial fans (and some deltas) formed (e.g., *2-3*). The fans mostly did not result from localized impact-induced precipitation and record a time span of >20 Myr of river flow (*2,11*). Data suggest that during this period, conditions were only intermittently wet enough for surface runoff (e.g., *4, 12-15*). The early period of valley networks and the later formation of alluvial fans show distinct spatial distributions of rivers (Fig. 1). (We exclude rivers not formed by precipitation, e.g., associated with groundwater outbursts, as well as other features that might record non-precipitation processes; Supplementary Materials). We use the shifts in the spatial distribution of precipitation-fed rivers (Fig. 1) to assess past changes in the strength of the atmospheric greenhouse effect. In broad outline, this approach has been attempted previously, using *Viking* data to analyze pre-Valley-Network-era changes (*16*). Here we study post-Valley-Network-era change using new data and new models, and draw a different conclusion.

**Results**

**Charting The Decline Of Mars' Surface Habitability.**

Given its potential as a probe of the ancient greenhouse effect, it is perhaps surprising that Mars' paleo-river distribution has not previously been analyzed for latitude/elevation trends with correction for detectability biases. We analyze river-distribution databases for >3.6 Ga Valley Networks (*17*), and for <3.6 Ga rivers (*18*) (Supplementary Materials). Both databases are corrected, for the first time, for post-fluvial resurfacing (i.e., detectability/preservation bias), masking out the areas shown in dark gray in Fig. 1 (the basis for this masking is explained in the Supplementary Materials).

To plot the elevations of <3.6 Ga rivers, we used the mean of the Mars Orbiter Laser Altimeter (MOLA) elevation of each of the eroded topographic catchments draining into observed fans and deltas. To plot the elevations of >3.6 Ga rivers, we used the topographic elevation of points midway along each (erosional) valley segment. Therefore, the comparison of the elevations of eroded pre-3.6 Ga landforms to the elevations of eroded post-3.6 Ga landforms is an apples-to-apples comparison (Fig. 2).

Contrary to uncorrected catalogs, Valley Networks formed preferentially at higher elevations than later-stage fluvial features (Fig. 2). At high elevations during the Valley Network era, there must have been a source of water (snow, ice, or rain), and in order to produce runoff the surface temperature, $T_{surf}$, must have exceeded 273 K, at least intermittently (e.g., *9, 19*). A very strong atmospheric greenhouse effect is required during the early (Valley-Network, >3.6 Ga)-era.

We neglect post-fluvial True Polar Wander (TPW) because it was likely minor (Supplementary Materials). Indirect geologic proxies have been used to infer post-Valley-Network TPW (*20*) but these are affected by resurfacing biases. If post-Valley-Network TPW was, in fact, ~20°, then the finding that Valley Networks formed at high elevation (Fig. 2) would be unaffected. Wiggling of the southern midlatitude band of <3.6 Ga rivers is well explained by coupling between atmospheric water transport and topography, without TPW (*21*).

**Global Climate Model.**

Surface liquid water requires a supply of $H_2O$ (e.g., snow or rain) and temperatures above freezing. Temperature would not be limiting if Mars had a very strong greenhouse effect, and surface liquid water would occur, at least seasonally, wherever $H_2O$ was available. If (as on Earth) the coldest parts of the surface act as a cold trap for snow and ice, then $H_2O$ would be available at locations where snow/ice is most stable, on a planetary surface that had only (3±2)% of Earth's area-averaged $H_2O$ abundance (*22*). By contrast, on a planet with a greenhouse effect that was weaker (but still stronger than on modern Mars), temperature would be limiting and liquid water could occur only in relatively warm locations.

To explore how changing patterns of surface temperature ($T_{surf}$) and snow/ice stability are driven by changes in $pCO_2$, non-$CO_2$ radiative forcing, and orbital forcing, we used the MarsWRF global climate model (GCM) (*23-24*) (Methods). $CO_2$ greenhouse warming is insufficient to explain pre-3.0 Ga rivers on Mars (*7*). Similar to refs. 25-26, our model represents the unknown, non-$CO_2$ greenhouse-warming agent by a gas with wavelength-independent "gray" opacity ($\kappa$) in the thermal infrared, with no opacity at solar wavelengths, and which does not contribute to total pressure. In previous modeling work investigating leading candidate agents of non-$CO_2$ greenhouse-warming, these agents contribute little to total pressure (probably <20%; ref. *19*), and are fairly well mixed horizontally; therefore, this representation is reasonable. $CO_2$ and the gray gas are the only species in the model. Solar luminosity is set to 80% of the modern value (*27*). Surface albedo and thermal inertia are set to uniform values of 0.2 and 250 J m$^{-2}$ K$^{-1}$ s$^{-½}$, respectively, which are close to Mars' average bare-ground values. The model is run to equilibrium and all output subsequently discussed is from the analysis of final-year output.

Our approach is to simply represent key physical processes in a fast-running model with relatively few adjustable parameters, including 3D topography as is necessary for comparison to paleo-river spatial distributions. This approach sits between that of 1D/2D models, and fully coupled GCMs. The most important limitation of our approach is that we do not simulate rainfall, although (as described below) we do test for the internal consistency and plausibility of meltwater runoff.

First, using $CO_2$-only runs, we re-verified (*9*) that at 1000 mbar $pCO_2$, snow is most stable on high ground, and the temperature dependence on latitude is, while still significant, relatively modest. Still higher atmospheric pressures would produce the same pattern (*9, 19*). A 20-mbar $pCO_2$ $CO_2$-only simulation is much colder, with strong latitudinal banding in $T_{surf}$ and snow stability. This is consistent with <1.5 Ga geologic data (e.g. *28*).

Next, we switched gray gas forcing on. For the strong non-$CO_2$ radiative forcing that is needed to explain geologic data, the surface temperature lapse rate (i.e., $\partial T_{surf}/\partial z_{topo}$, where $z_{topo}$ is the topographic elevation relative to the geoid) is approximately the atmospheric adiabatic lapse rate for all surface pressures. Greenhouse forcing, and not the density-dependent turbulent fluxes between the atmosphere and the ground, regulates the surface temperature lapse rate. Therefore, paleo-proxies for the atmospheric adiabatic lapse rate cannot be used as a paleo-barometer for steady-state wet climates on early Mars.

Finally, we ran 54 GCM simulations defining a 3×9×2 parameter space (Table S1), varying $pCO_2$ (20 mbar, 150 mbar, and 500 mbar); non-$CO_2$ radiative forcing (9 values log-evenly spaced from $\tau = 1.7$ to $\tau = 4.5$, where $\tau = \kappa \times (pCO_2 / 3.7$ m s$^{-2}$)); and orbital forcing (25° and 45° obliquity; obliquity is the most important orbital parameter for controlling Mars' climate). These parameters span a very wide range of greenhouse conditions, from globally sub-freezing up to warming strong enough for runoff everywhere on the planet. These parameters span the range of plausible parameters for early Mars river-forming climates.

Using the surface temperature and near-surface wind field output from the model, we calculated the likelihood of seasonal meltwater. First, we downscaled the 64×48 grid point GCM output to ~7 km-per-pixel using Mars Orbiter Laser Altimeter (MOLA) gridded topography. Temperature was downscaled using the run-by-run correlation between mean temperature and elevation for latitudes <30° and pressure was downscaled assuming a fixed scale height of 10.7 km. Following refs. 25 and 29, we assume that snow tends to accumulate in locations that minimize the annually-integrated sublimation rate. For example, on Earth, high ground acts as a cold trap for snow and ice (*9*). (Mars' topography has a range and standard deviation 3.5× greater than Earth's.) If so, cold traps contain snow and ice that (if temperatures get warm enough) becomes a water source for runoff during melt events (*9*). We represent this effect by a post-processing parameter, $f_{snow}$, corresponding to the areal fraction of the planet that has snow during the warm season. For a given value of $f_{snow}$, we check if any pixels with sublimation rates less than the threshold defining $f_{snow}$ have warm-season temperatures (mean temperatures for a continuous period of 100 sols) >273 K. The 100-sol requirement is to allow time for thermal maturation of snow/ice for runoff production (*30*). (This is conservative, and in the Antarctic Dry Valleys lake levels respond more quickly to temperatures above freezing; *31, 32*). Such pixels are designated as runoff pixels. We refer to this procedure as the Simplified Meltwater Model. Evaporitic cooling is not considered. For each GCM run, we repeat this for $f_{snow} = \{0.01, 0.02 \ldots 1\}$ to generate possible meltwater maps.

**Comparing Model Output To Geologic Data.**

We compare each map to the two geologic time slices (Fig. 1), assigning every (~7×7 km) pixel as a true positive (TP; runoff predicted, ≥1 water-worn features observed), true negative (TN; no runoff predicted nor water-worn featured observed), false positive (FP; runoff predicted, but no water-worn feature observed), or false negative (FN; no runoff predicted, but ≥1 water-worn features observed). We cosine-weight to correct for latitude. We marginalize over $f_{snow}$ by calculating (for each of the 54 GCM simulations), (1) a receiver-operating characteristic (ROC) curve varying $f_{snow}$ and measuring the area under the curve (AUC); (2) informedness (Youden's J statistic, defined as sensitivity + specificity – 1; Supplementary Materials); and (3) the precision-recall area-under-curve, PR-AUC). These calculations treat all values of $f_{snow}$ as equally likely, which represents our uncertainty about the Early Mars hydrologic cycle. Our procedure penalizes false positives and false negatives equally (Supplementary Materials). The three measures agree on which GCM run is the best fit for ~3.6 Ga, and the three measures agree on which GCM run is the best fit for <3.6 Ga.

For the Valley Network-era (>3.6 Ga), the model with the highest informedness (largest value of Youden's J statistic) (Fig. 4a; ROC AUC 0.68), corresponds to $\tau = 3.1$ ($\kappa = 5.7 \times 10^{-4}$), high (45°) obliquity, $f_{snow}$ = 0.43, $pCO_2$ = 150 mbar. Global annual-average temperature, $\overline{T}$, is 268 K (Fig. 3a). Seasonal meltwater runoff occurs over ~40% of the planet. Highland Valley Networks sit within the meltwater runoff zone. Meltwater runoff is absent at high latitudes and low elevations because warm-season snow is not stable there. (At 45° obliquity, high-latitude snow is destabilized by hot summer solstices; *33*). Still-warmer climates (e.g. *34*) – but not colder climates – also give good fits to the Valley Network data. (Global annual-average temperature $\overline{T} \geq 273$ K is required to explain all Valley Networks with a single forward model, at the cost of a slightly increased proportion of false positives.) In the best-fit model, annual-average temperatures above freezing exist over much of the Northern Lowlands, and in low spots including Gale crater, and Jezero crater (Figs. 3a, 4a). Many Valley Networks drain into these broad, no-permafrost zones.

For the fans-era (<3.6 Ga), the highest-informedness model (Fig. 4b; ROC AUC 0.65) corresponds to $\tau = 2.5$ ($\kappa = 4.4 \times 10^{-4}$), 45° obliquity, $f_{snow}$ = 0.55, $pCO_2$ = 150 mbar). This shows a much colder ($\overline{T} \sim 258$ K) planet. The highest ground is too cold for meltwater runoff (Fig. 3b), the low elevations (and high latitudes) are too warm for snow, and so meltwater runoff occurs at intermediate elevations, consistent with data. Notably, comparing the best-fit <3.6 Ga model to the best-fit 3.6 Ga model (Fig. 3), the changing spatial distribution of water flow on Mars does not require any change in $pCO_2$. All current rover landing sites, plus the *Rosalind Franklin* landing site, are predicted to have meltwater runoff production <3.6 Ga. Permafrost temperatures occur almost everywhere and so shallow-subsurface ice might protect surface water from loss to infiltration (at least temporarily). Infrequent spots with annual average temperatures above freezing, which would have unfrozen and therefore permeable subsurfaces, correspond to areas of known or suspected salt accumulation. This prediction is consistent with the groundwater-upwelling model for sulfate-deposit genesis (*35*). Geographically, the model performs well except for false negatives in North Arabia. Given the simplifications of our model, we cannot completely rule out an alternative option for the fans-forming climate that is warm and dry, but it is not favored by our model. A decisive future test is to use the *Curiosity* rover to search for meridianiite ($MgSO_4 \cdot 11H_2O$), which is unstable above 275 K.

How does the goodness of fit of the model to the data vary as a function of parameter values? Within the model, for both time slices, high obliquity is always favored over low obliquity. (By contrast, rainfall models permit low-obliquity river-forming climates; e.g., *36*). This is because, at low obliquity, the low latitudes (and especially high elevations at low latitude) are not particularly favored for meltwater runoff because snow is relatively unstable there; but that is where the valleys are. It remains possible that the patterns in Figs 1-2 are a palimpsest built up over a wide range of orbital forcings that cannot be represented by a run at a single obliquity value. This palimpsest hypothesis is one possible explanation for the valley networks in the southernmost highlands. Whether or not this hypothesis is true, high obliquity during the wet era is still favored. As high (45°) obliquities are favored for both time slices, changes in orbital forcing do not explain the changes between the two periods.

Figure 5 shows the trade-off between $\tau$, $pCO_2$, and model goodness. From this figure, we make the following observations.

1. Although the contours slope up and to the right, this slope is gentle. Thus, $\tau$ matters more than $pCO_2$. This is surprising, as regulation of $pCO_2$ is central to our understanding of planetary habitability (*7, 8, 37*). This result can be understood in terms of the high value of $\tau$ needed to match data. As $\tau$ is high in all infrared bands, the additional contribution to warming from the direct radiative effect of $CO_2$ is small, and counteracted by the increase in planetary albedo due to Rayleigh scattering. Of the runs warm enough to generate rivers, 150-mbar output has similar annual-average temperature to the 500-mbar output. Therefore decline in direct radiative forcing from $CO_2$ corresponding to $pCO_2$ decline from ~0.5 bar to ~0.15 bar was probably not the cause of the decline of Mars rivers. Within this $pCO_2$ range, where $T_{surf}$ depends only weakly on $pCO_2$, carbonate-silicate weathering feedback could have only a weak direct effect on climate. However, an indirect effect (through the effect of $pCO_2$ on the strength of $H_2$-$CO_2$ collision-induced absorption warming) remains possible. Because non-$CO_2$ radiative forcing is an increasing function of $pCO_2$ for some proposed non-$CO_2$ warming mechanisms (*38*), the door is still open to $CO_2$ decline being the ultimate driver of the decline of Mars rivers. Nevertheless, the focus shifts to the influence of non-$CO_2$ warming agents as a function of time.

2. It was previously proposed (*9*) that the Valley Networks' high-elevation preference indicates $\gg 10^2$ mbar paleo-$pCO_2$. However, Fig. 5 shows that these data can also be well-matched when $pCO_2$ is ~20 mbar.

3. The best fits for <3.6 Ga are ~10±5 K colder than the best fits for ~3.6 Ga. This makes sense given the downshift of ~5 km in zonally averaged river abundance (Fig. 2), assuming a surface temperature lapse rate of ~2.5 K/km. Zonal variations are not enough to confound this pattern. We infer that the lack of high-elevation <3.6 Ga rivers indicates a weaker total greenhouse effect at these later times, so that high-elevation precipitation stayed frozen. The drop in $\tau$ was likely bigger than Fig. 5b suggests, as solar brightening occurred during Mars' wet era but is not included in the model.

4. The $\tau$ for <3.6 Ga is well constrained (Fig. 5b). This corresponds to $\overline{T}$ ~ 258 K and a largely frozen Mars. The cold paleo-temperatures inferred from the spatial distribution of fans are consistent with other indications of snowmelt runoff, such as patchy erosion, the absence of channels on fans, and reports of aspect dependence (*18*). It remains to be seen if a rainfall model can match the data equally well.

5. The best fit for >3.6 Ga is $\overline{T}$ ~ 268 K and the upper limit is essentially unconstrained. This is consistent with the idea that Early Mars had $\overline{T} \geq 280$ K, with rainfall, especially if there were seas at low elevations (*31, 39*). On the other hand, if temperature was around the best-fit value and low elevations were dry, then most locations would have little or no rainfall.

In summary, we obtained paleotemperature estimates during the crucial interval when Mars was drying out (3.6-3.0 Ga and perhaps later). The result is consistent with estimates obtained using other methods. Overall, data-model comparison shows that changes over time in the distribution of Mars paleo-rivers in space (Figs. 1-2) are consistent with the expectation that Mars' greenhouse effect waned (perhaps non-monotonically) over time (Figs. 3-4). However, we did not find evidence from this method that the decline of surface habitability on Mars was associated with loss of atmospheric $CO_2$ (Fig. 5).

**Discussion**

Our forward model has limitations. Because warm-season temperatures and snow-covered areas are both derived from the same simulated year of GCM output, very rapid shifts in climate (too fast for snow/ice to relocate) cannot be included. This might contribute to the false negatives in northern Arabia Terra. Also, we do not include the albedo effects of snow (in effect, assuming dusty snow; *29, 30, 40*). Higher snow albedo would drive our best fits to higher $\overline{T}$ and set a correspondingly more stringent constraint on the strength of non-$CO_2$ warming needed to match data. It is possible that early Mars' atmosphere spent most of the time in a much colder state than shown in Fig. 5 (e.g., collapsed). In that case, our results still constrain the climate at times when Mars' climate supported rivers and lakes, so this is not a severe limitation.

To what extent are shifts in river distribution explained by the apparent three-fold decrease since 3.5 Ga in the inventory of Mars' surface/near-surface $H_2O$ (*41*)? $H_2O$ loss does not fully explain changes in river distribution. For example, groundwater table control does not explain the existence of mid-elevation rivers at 20-40° S (Fig. 1b). A weakening of the greenhouse effect is a simpler explanation for the data. The simplified meltwater model does not constrain the volume of surface/near-surface $H_2O$ (the $f_{snow}$ parameter describes the area of surface $H_2O$ cover, not $H_2O$ volume). The simplified meltwater model provides a simple description of the observed changes, but fully coupled global modeling of both the atmospheric and groundwater parts of the Early Mars water cycle has not yet been done. Therefore, determining whether $H_2O$ loss had a large or a small contribution to the observed changes remains a target for future work.

Our analysis suggests that the shifts in river distribution were driven by loss of non-$CO_2$ radiative forcing. The subsequent (<3 Ga; *3, 42*) cessation of river-forming climates on Mars could have been caused by further reduction in non-$CO_2$ greenhouse warming (perhaps the simplest explanation), by $H_2O$ loss, or by C loss (Fig. 6). The present-day rate of C escape-to-space is small (*43*), and isotopic evidence indicates that most of Mars' atmosphere was lost >3.5 Ga (e.g., *44-45*). Candidate carbon sinks include escape-to-space, carbonate formation, and basal melting of $CO_2$ ice. Alternatively, $CO_2$ could have been reversibly sequestered as $CO_2$ ice.

Major open questions remain, including the identity of the non-$CO_2$ greenhouse forcing agent that must have been in the atmosphere (at least intermittently) until ~3 Ga (Fig. 5). Although $H_2O$-vapor feedback would add to warming in all cases, it is insufficient (*9*). Hypotheses for the mechanisms for the needed additional warming include $H_2$-$CO_2$ collision-induced absorption (e.g. *38, 42-43*), or high-altitude water ice clouds (*48-49*). It is possible for either of these two mechanisms to provide very strong (stronger than needed to match data) warming, and this study provides new constraints on how much opacity is needed, and how this requirement changed over geologic time. These hypotheses can be tested using the *Perseverance* rover. *Perseverance* observation of detrital siderite (±fine-grained olivine) would favor moderate river pH and thus low p$CO_2$ during the years of river flow, potentially excluding warming by $H_2$-$CO_2$ collision-induced absorption. Similarly, *Perseverance* analysis of the texture and origin of carbonates (*50*) might help constrain pCO$_2$. Conversely, the global constraints on long-term evolution provided here offer long-term global context for inherently local (and likely stratigraphically restricted) sample acquisition by the Mars Sample Return program.

Mars is the only world whose surface is known to have become uninhabitable. This study quantitatively compared, for the first time, geologic data to a climate model over different eras within Mars' wet-to-dry transition. Future models of the evolution of <3.6 Ga river-forming climates should satisfy the geologic constraints on the shifting location of rivers (Fig. 2). We used a simplified meltwater model to perform the first (although not the only possible) analysis of the causes of this downshift in Martian habitability. Within our model framework, changes over time in the spatial pattern of paleo-rivers suggest a waning greenhouse effect. Comparison of a grid of GCM simulations combining gray gas and CO$_2$ radiative forcing shows little sensitivity of river-relevant output to pCO$_2$, but strong sensitivity to the strength of the non-CO$_2$ greenhouse forcing. Our results raise the possibilities that Mars had a thin (≤150 mbar) atmosphere throughout the river-forming era, and that Mars became uninhabitable as the result of constant-pressure cooling. This study does not rule out the hypothesis (*6-7*) that the end of surface habitability on Mars was ultimately driven by atmospheric decay, but suggest that loss of non-CO$_2$ radiative forcing (not loss of CO$_2$) played a dominant role in the changing spatial distribution of water flow from 3.6 to 3.0 Ga.

**Materials and Methods**

**Geologic data**
We used an existing Valley Network database (from 17), compiled using Thermal Emission Imaging System (THEMIS) data. Our new fans/deltas database (Fig. 1b) is described in ref. 18. This database was compiled by systematically searching between 90° N and 90° S using globally available Context Camera (CTX) images for fan-shaped deposits and their associated catchment areas. Details of how both databases were filtered and corrected for detectability bias are given in the Supplementary Materials.

**Description of global climate model simulations**
MarsWRF is a version of the PlanetWRF GCM, itself derived from the terrestrial mesoscale WRF model (*23-24*). Model resolution is 5.0×3.75 degrees (64×48 gridpoints) for longitude and latitude, respectively, with 40 model layers in height, spanning the surface to ~80 km. This resolution is typical for Mars GCMs. Further details are provided in the Supplementary Materials. For the gray gas, we use zero absorption coefficient for wavelengths less than 4.5 μm and an adjustable, constant absorption coefficient for wavelengths above 4.5 μm. The coefficient κ (m$^2$/kg) is chosen in order to produce a specified additional (non-CO$_2$) infrared optical depth, τ, at reference elevation (0 m). GCMs are run until the annual cycle has converged (typically <10 simulated years), at fixed orbital conditions. To get from the GCM model output to predicted meltwater runoff locations, we used a simplified meltwater model. The model assigns snow/ice to pixels that are calculated to have relatively low sublimation rate. Then, the model outputs a melt prediction for pixels where the temperature is warm enough for runoff. Further details, including how the predictions from the different GCMs were scored in comparison to the data, are given in the Supplementary Materials.

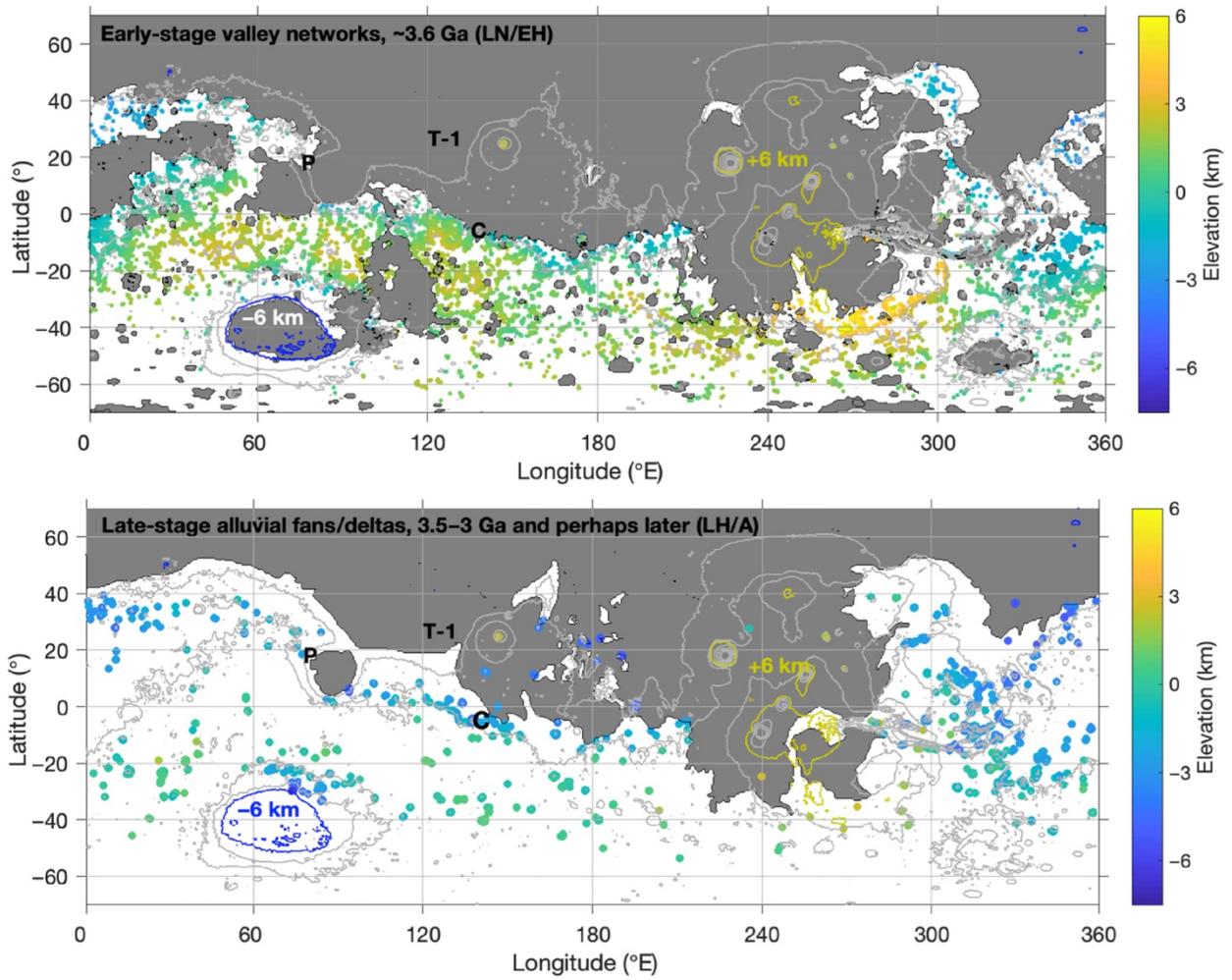

**Fig. 1. Changing spatial distribution of water flow on Mars.** *Top:* Distribution of >3.6 Ga (Late Noachian / Early Hesperian, LN/EH) rivers (*17*). Color: river elevation. Gray: excluded region (low/no detection probability). Rivers plotting within gray region are shown by black dots. Rovers: C=*Curiosity* at Gale crater, P=*Perseverance* at Jezero crater, T-1=*Tianwen-1* rover (*Zhurong*). Elevation contour spacing is 3 km. *Bottom:* Distribution of <3.6 Ga (Late Hesperian / Amazonian, LH/A) rivers (*18*).
9

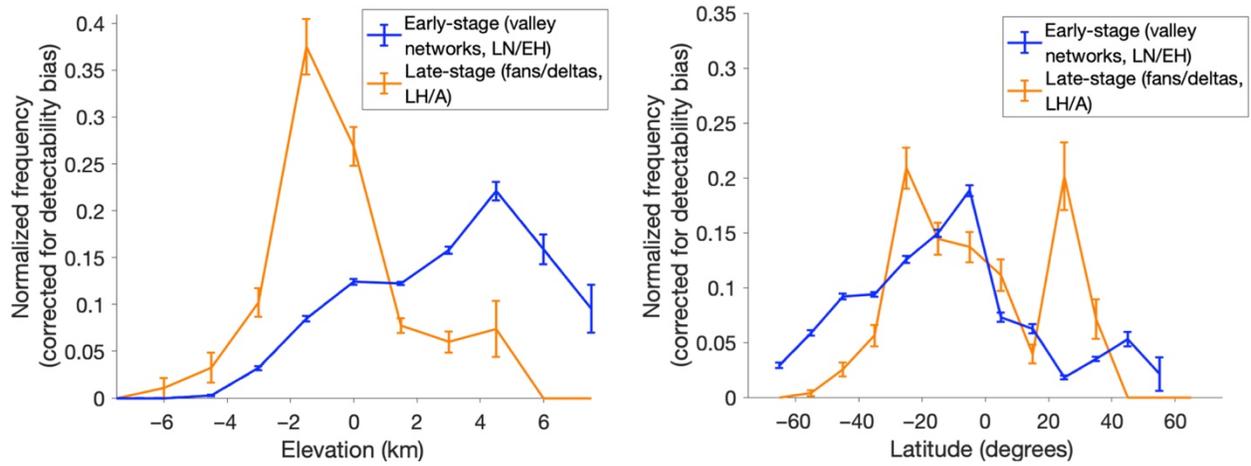

**Fig. 2. Decline over time in elevation of water-worked landforms, corresponding to a shift from >3.6 Ga elevation control to <3.6 Ga latitudinal banding.** Normalized frequencies (±1σ) of the elevation (*left*) and latitude (*right*) distribution of fluvial features on early Mars (valley networks) and later Mars (fans/deltas). Figs. S3-S5 give more details and sensitivity tests.

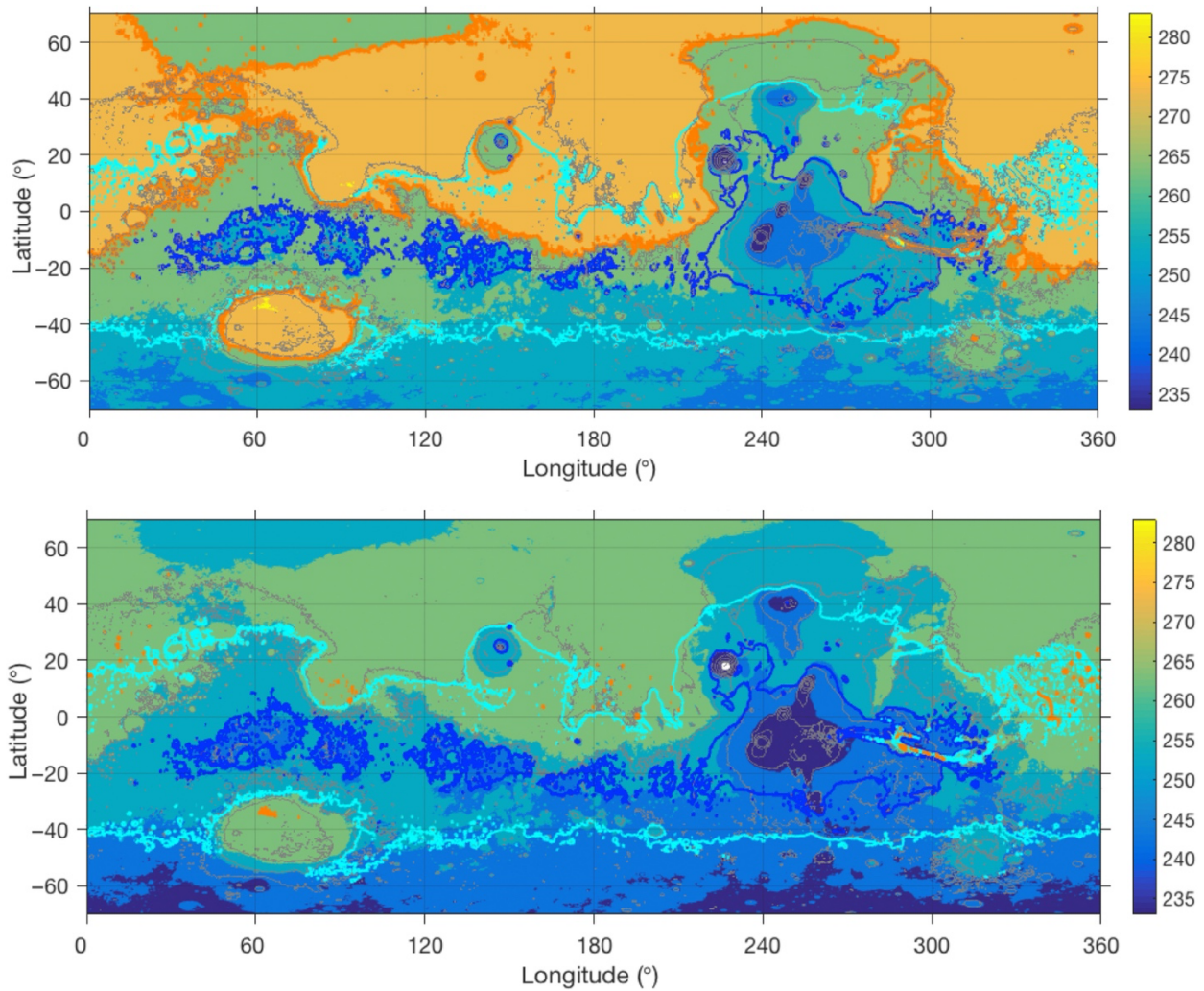

**Fig. 3. Global Climate Model output.** (**a**) Annual-average output for GCM run that best matches ~3.6 Ga data ($\tau = 3.14$, 45° obliquity, $pCO_2 = 150$ mbar). Color shading corresponds to annual-average temperature, (K). Orange line highlights 273 K isotherm and color bands mark 10 K intervals. Dark blue and cyan contours outline cold traps. The dark blue contours contain the part of the planet's surface area with the lowest annually integrated snow sublimation rate (0-10th percentile), and the cyan contours contain the part of the planet's surface area with lower-than-average annually integrated snow sublimation rate (0-50th percentile). Output (64 × 48) is downscaled using MOLA topography. Elevation contours (gray) are spaced at 3 km intervals. (**b**) As (a), but for the GCM run that best matches 3.5-3.0 Ga data ($\tau = 2.46$, 45° obliquity, $pCO_2 = 150$ mbar).

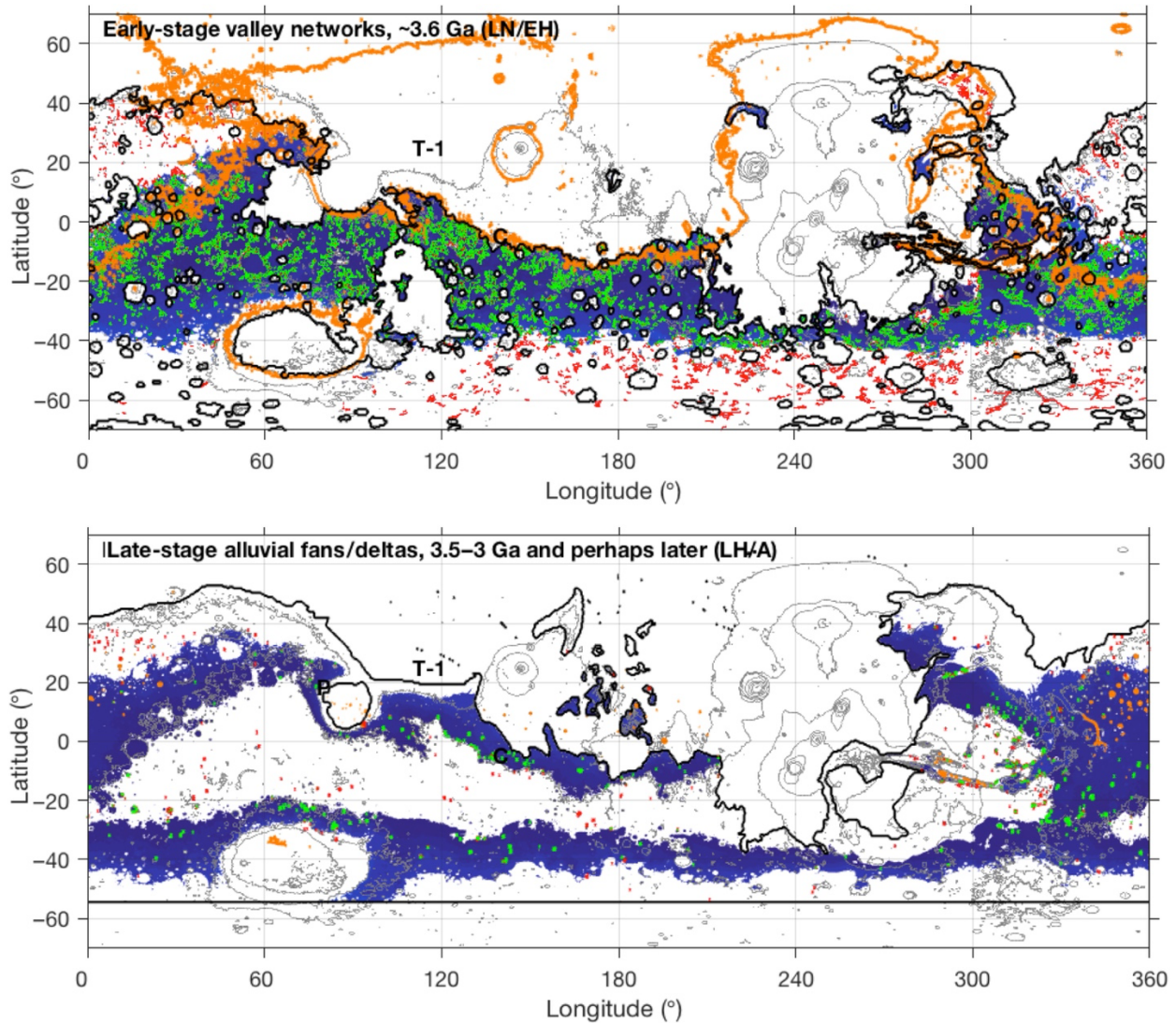

**Fig. 4. Comparison of data to model predictions. (a)** Data-model comparison for simplified snowmelt model output that best matches ~3.6 Ga data ($\tau = 3.1$, 45° obliquity, $pCO_2 = 150$ mbar, $f_{snow} = 43\%$). Dark blue band corresponds to area of snowpack with a continuous 100-sol period with average temperature exceeding 273 K, corresponding to predicted snowmelt runoff. Green symbols correspond to true positives (mapped water-worn features within the predicted snowmelt runoff zone). Red symbols correspond to false negatives (mapped water-worn features outside the predicted snowmelt runoff zone). The thick black line outlines the masked-out region (low/no detection probability for water-worn features), which differs between the two eras. Uncommon gray symbols show water-worn features within the masked-out area. Elevation contours (gray) are spaced at 3 km intervals. The orange line highlights the 273 K isotherm in annual-average temperature. (Youden's J, 0.30; ROC AUC for varying $f_{snow}$ for this GCM, 0.68). Rovers: C=*Curiosity*, P=*Perseverance*, T-1=*Tianwen-1* rover (*Zhurong*). **(b)** As (a), but for the simplified snowmelt model output that best matches 3.5-3.0 Ga data ($\tau = 2.46$, 45° obliquity, $pCO_2 = 150$ mbar, $f_{snow} = 56\%$). (Youden's J, 0.34; ROC AUC for varying $f_{snow}$ for this GCM, 0.65).

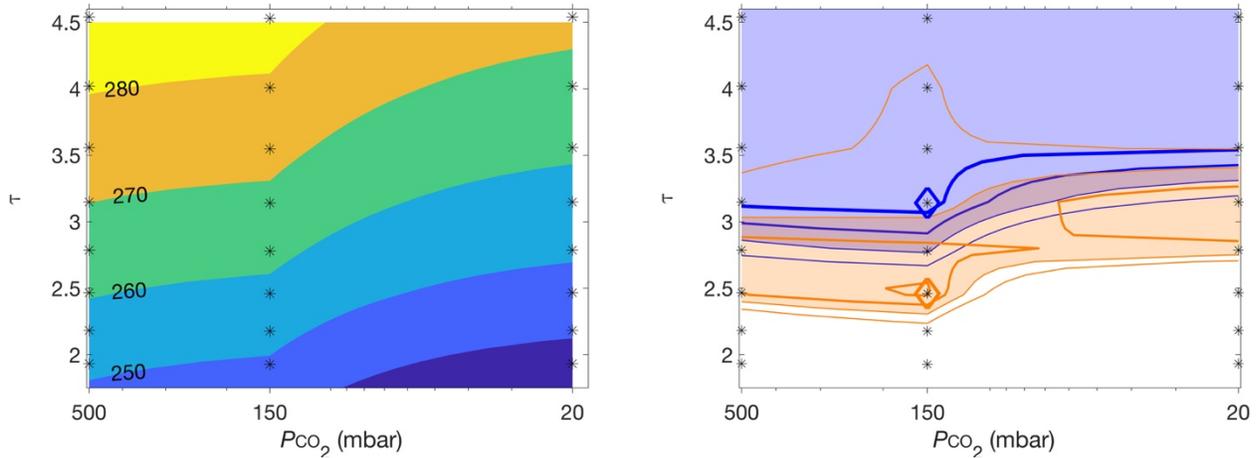

**Fig. 5. Best-fit Mars climate versus time. (a)** Global annual average temperature (K) as a function of $pCO_2$ and of gray gas column optical depth, $\tau$. Black asterisks correspond to inputs to individual GCMs, and values between the asterisks are interpolated. Obliquity = 45°. **(b)** Goodness of fit of model to data as a function of $pCO_2$ and of gray gas column optical depth, $\tau$. Blue shaded region corresponds to relatively good fit to >3.6 Ga data, and orange shaded region corresponds to relatively good fit to <3.6 Ga data. The blue diamond is the best-fitting GCM run for >3.6 Ga data, and the orange diamond is the best-fitting GCM run for <3.6 Ga data. Metric is ROC AUC; contours correspond to 0.5, 0.55, 0.6, and 0.65 (thicker lines correspond to better fit). Asterisks correspond to inputs to individual GCMs, and values between the dots are interpolated. Figs. S14-S15 give more details and sensitivity tests.

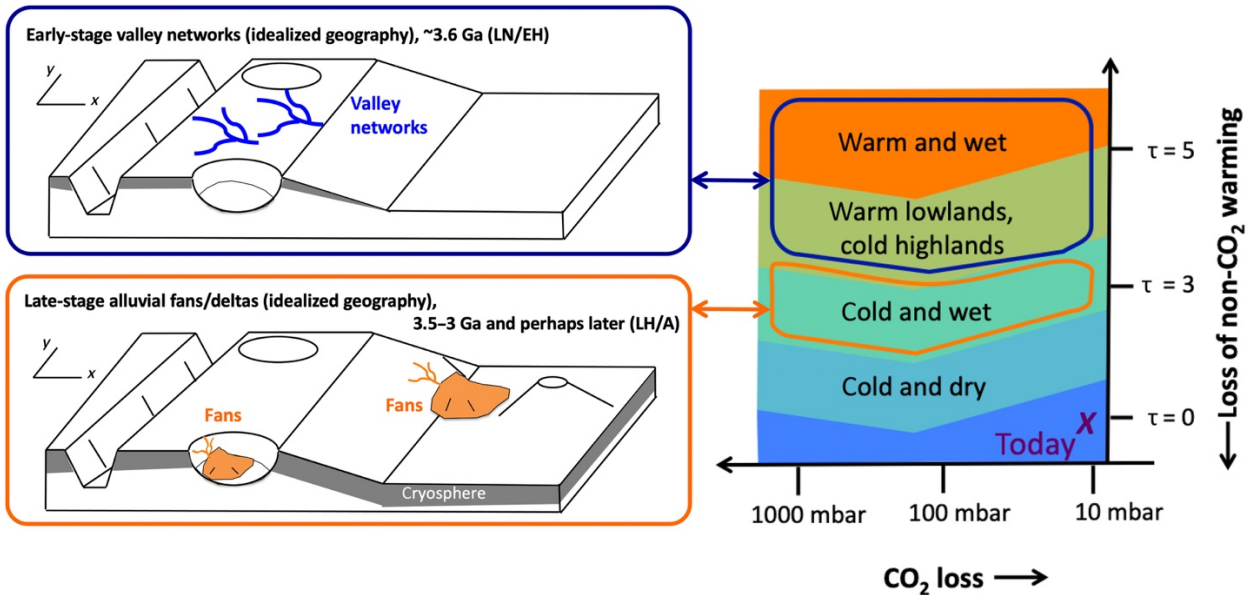

**Fig. 6. Graphical summary of this study.** (a) Left column shows a simple (geographically idealized) schematic of changes over time, interpreted by analysis of elevation decline (Figs. 1-2) in comparison to the ensemble of GCMs (Figs. 3-4). Right column: cartoon summary of Fig. 5, illustrating that within the framework of our model, shifts can occur with or without changes in $pCO_2$, but very probably require a decline in non-$CO_2$ radiative forcing.

**Acknowledgments**
We thank B. Hynek, D.P. Mayer, A.O. Warren, N. Dauphas, J. Dickson, G. Stucky de Quay, R. Ramirez, D. Jablonski, HiWish, and the University of Chicago Research Computing Center (RCC). We thank five anonymous reviewers for timely, accurate, and useful comments.

**Funding:** NASA (80NSSC20K0144+80NSSC18K1476). A portion of this work was performed at the Jet Propulsion Laboratory, California Institute of Technology, under contract with NASA. Resources supporting this work were provided by the NASA High-End Computing (HEC) Program through the NASA Advanced Supercomputing (NAS) Division at Ames Research Center.

**Author contributions:** E.S.K. conceived the work, analyzed the data, carried out the GCM runs, and wrote the manuscript. M.A.M. and B.F. contributed to project design, GCM modeling, and interpretation. A.M.M. and S.A.W. carried out the survey of alluvial fans/deltas. M.I.R. supported the GCM modeling.

**Competing interests:** Authors declare that they have no competing interests.

**Data/materials availability:** All data needed to evaluate the conclusions in the paper are present in the paper and/or the Supplementary Materials. Additional data are stored at Zenodo (doi:10.5281/zenodo.6380945). The MarsWRF source code can be made available by Aeolis Research pending scientific review and a completed Rules of the Road agreement. Requests for the MarsWRF source code should be submitted to: Mark I. Richardson (mir@aeolisresearch.com).

# Supplementary Materials

**Supplementary Text**

**1. Geologic background.**

Early Mars had rivers and lakes (e.g., *1*, and references therein) (Fig. S1) that were filled with liquid water (*51*). In this study, we use rivers that were fed by precipitation (rain and/or meltwater) (e.g., *4,* and references therein). On Mars, the channel heads for young rivers can, in many locations, be traced up to ridgelines (e.g., at 24.1°S 28.2°E, 22.7°S 73.8°E, 21.4°S 67.3°E, and 19.9°S 327.3°E). This is as expected for precipitation-fed runoff but inconsistent with spring discharge. Moreover, drainage density at some sites is much higher than expected for spring-fed streams, and it is difficult for spring-fed streams to incise into bedrock as is required to explain the Mars data (*52*). We do not distinguish between shallow subsurface flow (within topographic catchments) and overland flow. Both are fed by precipitation (rain or snowmelt), and so have similar implications for paleoclimate.

We omit from this study channels that were formed by catastrophic outbursts of subsurface water and we omit almost all features where there are arguments both for and against a groundwater origin (e.g., *53-54*). We do interpret as precipitation-fed a small number of deltas that were interpreted by ref. *55* as formed by groundwater discharge. However, because our results are robust to excluding all deltas, this difference in interpretation is not important for our conclusions.

Multiple lines of evidence show that river water on Mars 3.5-3.0 Ga was mostly not produced by the localized effects of asteroid impacts (e.g., *2, 11, 18, 56)*. For example, river/lake sediments found within large impact craters frequently contain smaller prefluvial/synfluvial impact craters (*2*). In order for these craters to accumulate, a time interval between the formation of the host craters and the end of river activity of ≥0.2 Gyr is required (e.g., *11*). The lifetime of post-impact hydrothermal activity is very much shorter, precluding post-impact hydrothermal activity as the source for river water. Moreover, simulations of impact-induced precipitation predict ≲1 yr-duration wet climates, too brief to match data for <3.6 Ga rivers and lakes (e.g., *56*, and references therein). The energy source for surface runoff was not impact energy, but insolation, based on the N-S orientation preference that has been reported for <3.6 Ga fans (*18*). Insolation-controlled (not impact-controlled) alluvial fan formation is also supported by the strong latitude dependence of fan-hosting craters (Fig. 2). The alluvial fan deposits are up to 1.1 km thick (*11*), which is, at best, only marginally consistent with a direct impact trigger for fan-forming rivers. It remains possible that impacts indirectly triggered wet climates by activating a mechanism that trapped sunlight energy (for example, through low-albedo impact ejecta, or by activating a hysteresis in the climate system, or by releasing $H_2$, *57*). Some steep (>10°) young (<1 Gya?) fans that are too small (<10 $km^2$, usually <1 $km^2$) to be included in the Fig. 1b survey probably did result from the localized effect of asteroid impacts (e.g. *58*). Possible water sources for these fans include localized precipitation or fall-back of wet ejecta. Moreover, some asteroid impacts on Mars triggered melting or release of subsurface fluids (e.g., *59*), but these are easily distinguished from landforms formed by precipitation-fed rivers. We also omit from our survey young gullies, which might be formed by mass wasting (for example, of $CO_2$ ice; *60*).

Intense geothermal heating (*61*), or lava, can melt ice, but high runoff production rates (e.g., *62*) rule out geothermal meltwater as the source of runoff. Lava deposits are not found in the locations where valley networks and paleolakes are seen, with rare exceptions that are excluded from our databases.

Chronologic constraints: The >3.6 Ga Valley Networks are distinguished from <3.6 Ga rivers on the basis of crosscutting relationships, crater densities, and the degree of preservation of crater rims (e.g., *3*, *42*). The <3.6 Ga rivers were usually shorter, with less integration of drainage basins (*63*), and <3.6 Ga crater rim erosion is more concentrated into alcoves (e.g., *64*). Later periods of river-forming climate show spatially more restricted fluvial erosion, (i.e., shorter rivers), and fewer aqueous minerals, but with high rates of peak runoff production (*3*, *51*, *62*). In many places, an epoch of deep wind erosion stratigraphically separates >3.6 Ga rivers from <3.6 Ga rivers (e.g. *14*). On Mars, almost all mappable-from-orbit fans/deltas postdate the Valley Networks. Because of these and other differences it is generally straightforward to distinguish between Valley Networks and <3.6 Ga rivers. However, detailed study of one Valley Network suggests that it was incompletely reactivated during later wet events (*65*). The <3.6 Ga rivers date from the Late Hesperian and Amazonian (e.g., *3, 38*). The currently most widely used mapping between relative and absolute age (*66*), which is uncertain by up to ±1 Gyr, is Noachian epoch → pre-3.6 Ga, Hesperian epoch → 3.6-3.2 Ga, Amazonian epoch → post-3.2 Ga. Ref. *10* reports crater counts on Valley Networks that are consistent with Valley Network formation during a <0.2 Ga interval around the Noachian-Hesperian boundary.

## 2. Detectability-corrected surveys for presence/absence of rivers.

We used an existing Valley Network database (from *17*). Other global Valley Network databases compiled using other methods show the same Valley Network spatial distribution pattern, with only minor differences (e.g., *67*). Most Valley Networks formed >3.6 Ga (*10*); a small percentage of Valley Networks are young and those were excluded from our analysis using criteria described below.

Our new fans/deltas database (Fig. 1b) is described in ref. 18. Fan-shaped deposits were classified, based on the presence of an arcuate frontal scarp indicative of subaqueous deposition, as being either putative deltas or alluvial fans. The fan apex was marked as the location where MOLA-derived contours indicated a transition from convergent to divergent flow, and the fan toe defined as the minimum elevation along a profile that extends from the apex to the point that is furthest from the apex along the fan outline. Fan and catchment areas were mapped using CTX images and MOLA-derived gridded elevation data. Elevation values for both the fan and the catchment were obtained from the MOLA 128 pixel/degree Digital Elevation Model (DEM). In total, our fans/deltas database has 1468 fans and deltas, 392 outside craters and 1076 inside craters (with 320 unique fan and/or delta-hosting craters). The largest previous survey involved ≈65 fans (*68*). We excluded 110 "terraced" deltas from further analysis. This is because terraced fans have been hypothesized to result from rapid groundwater outbursts (*53*). We also excluded 560 fans/deltas with fan area < 10 km$^2$ for the reasons described above. With these exclusions, 854 fans and deltas remained. These exclusions turn out to not make a big difference to the results (Fig. S3c).

We corrected both databases for detectability bias (post-fluvial resurfacing). A geologic feature of a given relative age has low/no detection probability if the terrain was subsequently resurfaced by a geologic unit of a younger relative age. Relative-age assignments were made using the United States Geologic Survey (USGS) Global Geologic Map of Mars (*69*) and also a lower-resolution map by ref. *70* on a 2-pixel-per-degree grid. For the Valley Network-era time slice, terrain not mapped as "Noachian" was masked out. Terrain mapped as "HN" (Hesperian/Noachian), e.g., Meridiani Planum, was also masked out. A small percentage of Valley Networks postdate ~3.6 Ga (*10, 17*) – these usually have fewer branches and are usually superposed on young volcanoes. In order to remove these contaminants, we used age assignments mapped by ref. *10* as well as more recent geologic mapping (*69*) to remove valleys that crosscut (and so postdate) Hesperian or Amazonian terrain. We removed Valley Networks in the following zones: 45-86°E / 45-65°S (Malea Planum), -150 to -86°E / 0-50°N (Alba Patera and high Tharsis), 120-180°E / 11-45°N (Elysium), and 75-97°E / 30-50°S (E. Hellas). The combination of these masking steps reduced the number of valley segments from 59419 to 53489. (For legibility, in Fig. 1a we only plot every 4$^{th}$ valley segment). Black dots in Fig. 1a show segments inside our masked-out zone. Most valley segments are part of a larger valley network ($n$ = 9879, of which 9278 survive our masking). For Fig. 2, when calculating error bars, we used the $\sqrt{N}$ error associated with the number of valley networks. For each valley segment, the location of a point roughly halfway along that segment (mean segment length = 12 km) was taken to be representative of the latitude and longitude of the segment. Using these coordinates, valley segment elevation was assigned using 8-pixel-per-degree MOLA data (*71*). Valley Networks are underrepresented in Arabia Terra due to erosion, burial, and the inverted-relief preservation style of fluvial materials in this region (e.g., *17*, *72*). Therefore, we excluded from our Valley Network analysis the region where abundant inverted river deposits are found, but few or no river valleys are found (*72*) (Fig. S8). Our conclusions turn out to be unaffected by this decision (Fig. S4). For the analysis of <3.6 Ga features, we masked out only terrain mapped by ref. *70* as either "Early Amazonian Vastitas Borealis Unit" or as "Late Hesperian – Late Amazonian volcanic materials". Although our post-fluvial resurfacing correction corrects for most of the detectability bias in our dataset, some post-fluvial processes can compromise the detection of fluvial features without fully resurfacing the landscape. Specifically, geologic features poleward of 40° in both hemispheres can be obscured by Amazonian slope-softening processes (*73*), and terrain at < -6 km elevation is on the Hellas basin floor, which has enigmatic geology. Our results are robust to the exclusion of these areas.

After masking, in order to make Fig. 2, the non-masked area (km$^2$) falling into every 2D elevation+latitude bin was calculated (Fig. S5). Also, the number of the corresponding geologic features falling into each 2D bin was calculated. In the case of Valley Network segments, segment frequencies were weighted by valley segment length. The outputs were marginalized on elevation or latitude. Then, the ratio of feature frequency to non-masked terrain area was taken to calculate relative frequency (units of km/km$^2$ for Valley Networks, and features/km$^2$ for fans). The resulting relative frequency histograms were normalized such that the area under each curve was unity (Fig. 2). Our inference of latitude-dependent Valley Network formation (Fig. 2) is supported by independent measurements of valley depth (*74*). Mean valley depth declines from ~140 m within 10° of the equator to ~110 m further (10-30°) from the equator (*74*).

The detection of a decline in the elevation of rivers over geologic time is robust to different choices about which elevations to plot. Sensitivity tests for different elevation-assignment choices are shown in Figs. S2-S3. A decline in the elevation of rivers over geologic time is seen regardless of whether we use fan apex elevation, fan toe elevation, or maximum catchment elevation for fans/deltas, and regardless of whether we set the minimum Strahler order for Valley Networks to 1 or 3 (Fig. S4).

Post-fluvial tectonics was likely minor: Mars lacks plate tectonics and the amplitude of post-valley-networks True Polar Wander (TPW) was insufficient to produce the faulting pattern predicted for TPW (75). <3.6 Ga rivers show a distribution that parallels latitude bands, inconsistent with large-amplitude TPW after the <3.6 Ga rivers (Fig. 1b). Post-Valley-Network resurfacing (gray area in Fig. 1a) is the main cause for the >3.6 Ga Valley Networks showing a distribution that is roughly sinusoidal with longitude. A small-circle-fitting algorithm was applied by ref. 20 to the distribution of Valley Networks, and a paleopole at 69°N 118°W was proposed; however, this workflow neglected resurfacing. We applied this algorithm to a synthetic dataset in which, by construction, there is no polar wander, and found roughly the same paleopole (Fig. S7). Specifically, ref. 20 fits a small circle to the observed distribution of Valley Networks, assuming that the Valley Networks originally formed in a latitude belt that has been distorted by TPW into the observed roughly-sinusoidal shape (Fig. 1a). We applied the same TPW-retrieval procedure to synthetic data in which the distribution of Valley Networks is spatially uniform – i.e., by construction the synthetic data contain zero evidence for TPW (Fig. S7). Before running the paleopole-retrieval algorithm, we crudely implemented post-fluvial resurfacing by removing valleys in areas greatly affected by glacial resurfacing (poleward of 55°), in Arabia Terra (diagonal cut in center of Fig. S7), and in tectonized areas in Acheron and Tempe Terra (affected by faults such as Mareotis and Tempe Fossae). Ref. 20 (their Figure 1) does not show any valleys in Xanthe Terra, or SE of Argyre. For consistency with Figure 1 in (20), we also removed valleys in these regions, even though Noachian valleys in Xanthe Terra and SE of Argyre are mapped in ref. 17 (their Figure 1). The paleopole obtained by fitting a small circle to the synthetic data surviving this masking procedure is 79°N 122°W. The Valley Network-based paleopole reported by ref. 20 is 69°N 118°W (Fig. S7). These values are similar (Fig. S7), demonstrating the importance of correcting for subsequent resurfacing before application of the small-circle-fitting procedure of ref. 20 to Valley Network-era geologic data. Ref. 20 does not make such a correction, and on the basis of our calculation we believe that such a correction is a necessary step before making the claim that the Valley Network distribution is evidence for TPW on Mars (20).

Although it was once thought that the deviation of putative Mars Northern Ocean shorelines from an equipotential might be evidence for TPW, improved calculations from the same research group show that "Tharsis-induced TPW has a negligible effect" on the deviation of putative Mars Northern Ocean shorelines from an equipotential (76). Moreover, putative Mars Northern Ocean shoreline features have been questioned based on high-resolution data. If half or more of Tharsis was emplaced before the Valley Networks, then the theoretical upper limit on Tharsis-induced TPW is small, <9° (76). There is no need to appeal to TPW to explain the local lack of Valley Networks in E. Noachis (around 40°E 30°S) because this lack can be explained as due to a standing wave in the atmospheric circulation forced by ancient topography (e.g., 21, and references therein). If large-amplitude (~20°) post-Valley Network TPW did, in fact, occur on Mars, our conclusion that Valley Networks formed at high elevation (Fig. 2) would be unaffected.

Light-toned, layered sedimentary rocks: Another geologic proxy for past near-surface liquid water that also postdates the Valley Networks is the distribution of light-toned layered sedimentary rocks (29, 77). The water for aqueous cementation of these materials could have been supplied in the context of diagenesis by groundwaters, not necessarily surface flows (e.g., 14). This proxy shows a low-elevation preference and strong latitude dependence, consistent with the shifting controls on river distribution. Fig. S6 shows the distribution of light-toned, layered sedimentary rocks as cataloged by ref. *77*. For a more inclusive catalog showing a broader distribution of sedimentary rocks, see *78*.

## 3. Global climate model (GCM) simulations.

Description of global climate model simulations.
In order to explore how patterns of temperature and snowpack stability change as a function of $pCO_2$ and of gray gas forcing, we used the MarsWRF GCM (*24*). The vertical grid uses a modified-σ (terrain-following) coordinate system with 40 vertical resolution levels. Periodic boundary conditions in the horizontal dimensions are employed, and an absorbing ("sponge") upper boundary condition was used. $CO_2$ radiative transfer uses a correlated-$k$ scheme (*79*). A central aim of this study is to decouple the effect of loss of $CO_2$ from the effect of non-$CO_2$ greenhouse warming. To represent the non-$CO_2$ greenhouse warming, we use a gray gas approach (e.g., *25-26*). A well-mixed gray gas with adjustable absorption coefficient can approximate the effect of many potential non-$CO_2$ Mars warming agents (e.g., $H_2$, $CH_4$). For example, although cloud coverage on Earth is far from complete, water ice cirrus clouds can only provide strong greenhouse warming on Mars if cloud coverage is close to complete (e.g., *80*).

The dynamical time step varies between simulations (in the range 30 s – 180 s). The surface layer flux parameterization scheme is a Monin-Obukhov scheme. We assume $pCO_2$ is approximately equal to total atmospheric pressure; Mars's atmosphere is 95% $CO_2$ today, and atmosphere evolution calculations indicate that $CO_2$ was also the dominant constituent of the atmosphere in the past. Total atmospheric pressure would increase by <20% due to leading candidate agents of non-$CO_2$ greenhouse warming ($H_2$, $CH_4$, and high-altitude water ice clouds). We allow $CO_2$ ice clouds to form and we permit $CO_2$ seasonal polar cap condensation. We imposed: present-day topography (from MOLA; *71*), spatially uniform surface albedo (0.2) except when ($CO_2$) ice is on the surface, spatially uniform surface thermal inertia (250 J m$^{-2}$ K$^{-1}$ s$^{-½}$), and no atmospheric water cycle nor atmospheric dust. "Dust continents" (high-elevation, high-albedo, low-thermal-inertia zones) are omitted because their formation age is unknown and these "dust continent" features might postdate all of the fluvial features in our databases. Mars's geologic record of river forming climates spans 2 Gyr, and almost certainly samples multiple orbital states due to chaotic obliquity variations (e.g., *33*). Therefore, we carried out all our runs at both low (25°) and high (45°) obliquity. Eccentricity was set to zero, and all calculations were performed at 80% of present-day solar luminosity (as appropriate for ~3 Gya).

GCMs are run until the annual cycle has converged (typically <10 simulated years), at fixed orbital conditions. Output was sampled 260 times per year, allowing good sampling of the diurnal cycle in each season while preventing time-of-day aliasing (Fig. S9). Future studies might include more frequent output to fully resolve each diurnal temperature cycle, allowing direct computation of degree-days above freezing and meltwater production. The potential sublimation rate, $S_{pot}$, is defined as the sublimation that would be experienced by snow/ice if that snow/ice were forced to have the same temperature as simulated for bare ground; it is a measure of relative snowpack stability (low potential sublimation rate = greater likelihood of snow accumulation; *25, 29*). Details of how this is calculated are provided later in this Supplementary.

To check the precision of our model, we analyzed the energy conservation properties of our runs. In steady-state, ignoring energy uptake and release by the ground, annually-integrated and planet-averaged absorbed solar radiation equals annually-integrated outgoing longwave radiation. Averaging over the last year of the model run, the best-fit 20-mbar runs have an energy imbalance of <2.5 W/m², the best-fit 500-mbar runs have an energy imbalance of 5.1-6.2 W/m², and the best-fit 150-mbar runs (which are also the best overall fits shown in Figs. 3-4) have an energy imbalance of 2.3-3.0 W/m². This is sufficient for the purposes of our study.

The simplified meltwater model.

To get from the GCM model output to predicted meltwater runoff locations, we used a simplified meltwater model. The model assigns snow/ice to pixels that are calculated to have relatively low sublimation rate. Then, the model outputs a melt prediction for pixels where the temperature is warm enough for runoff (guided by terrestrial experience and ref. *30*, we choose a continuous 100-sol period with an average temperature >273 K for this threshold). The approach has been used before (*29*) and is consistent with Early Mars GCM output (*25*). The major advantages of the simplified meltwater model are: (i) hydrologic-cycle uncertainties are collapsed into a single parameter, $f_{snow}$ (the fraction of the planet that has warm-season snow; small $f_{snow}$ corresponds to patchy snow, and large $f_{snow}$ corresponds to very extensive snow), (ii) fast runtime, and (iii) ease of downsampling to the smaller-than-GCM-pixel scale of the features of interest. When used in combination with the parameters used for the GCM-run ensemble, the simplified meltwater model also has restrictions and limitations: by construction, the model does not explicitly simulate rainfall (*19*, and references therein), we are not considering variations in orbital eccentricity, mesoscale processes such as orographic precipitation and steep slopes, nor the increase in solar luminosity during the period of river-forming climates on Mars. The flat albedo means that we are effectively assuming dusty snow (*40*).

Pressure, temperature $T_{surf}$, and wind speed $|u|$ for each of the 260 time steps in final-year model output are downsampled onto an 8 ppd topographic grid based on Mars Orbiter Laser Altimeter topography. In order to do this, the temperature is corrected for the adiabatic lapse rate (Fig. S12). The pressure is assumed to be a function of topography (a constant scale height of 10.7 km is assumed), but not time or latitude. The lapse rate is calculated by linear regression of the average temperature on the topography for latitudes <30°. The corresponding lapse-rate-based temperature offset is applied uniformly to get downsampled temperature as a function of time.

The warm season temperature at each of 2880 × 1440 pixels, $T_{warmseason}$, is calculated by finding the warmest 100-sol-average temperature (with a cyclic wrap). We assume $T_{warmseason}$ > 273 K is required for runoff. This is a more stringent constraint than just producing some meltwater at the surface, which is appropriate because runoff requires thermal maturation of the snowpack to prevent refreezing during infiltration (*30*), and this choice is conservative.

The potential sublimation, $S_{pot}$, is calculated in post-processing as the sum of the potential forced sublimation and the potential free sublimation. The maximum annually integrated sublimation potential was calculated using the bare-ground temperature, combining the forced turbulent flux and free turbulent flux, assuming planetary boundary layer relative humidity of 50%. The potential free sublimation, $S_{free}$, (kg/m²/s) is calculated as in refs *29* and *81*:

$$D_a = 1.387 \times 10^{-5} (T_{surf} / 273.15)^{1.5} (10^5/P) \qquad (1)$$

$$v_a = 1.48 \times 10^{-5} (8.314\ T_{surf}/(0.044\ P))\ ((240 + 293.15)./(240 + T_{surf}))\ (T_{surf}/293.15)^{1.5} \quad (2)$$

$$e_{sat} = 611.2\ \exp(2834000 / 461.5\ (1/273.16 - 1/T_{surf})) \quad (3)$$

$$\rho_{sat} = e_{sat}/(R_{H2O}\ T_{surf}) \quad (4)$$

$$\rho_a = P / (8.314 / 0.044\ T_{surf}) \quad (5)$$

$$\Delta\eta = \rho_{sat}\ (1-0.5) / \rho_a \quad (6)$$

$$\Delta\rho/\rho = (0.044-0.018)\ e_{sat}\ (1-0.5) / ((0.044\ P)) \quad (7)$$

$$S_{free} = 1.57 \times 0.14 \times \Delta\eta\ \rho_a\ D_a\ ((v_a./D_a)\ (g/v_a^2)\ (\Delta\rho/\rho))^{1/3} \quad (8)$$

where $D_a$ is diffusivity, $v_a$ is kinematic viscosity, $R_{H2O}$ is the specific gas constant for water, $g = 3.7$ m/s$^2$ is Mars gravity, and we have assumed that the boundary layer temperature, $T_{bl} \approx T_{surf}$. The forced potential sublimation $S_{forced}$ (kg/m$^2$/s) is calculated as

$$S_{forced} = C_D\ |u|\ e_{sat}\ (1-0.5) /(R_{H2O}\ T_{surf}) \quad (9)$$

where $C_D$ is the drag coefficient, set to a fixed value of 0.00275 (following ref. *25*). (Note that interpolation of wind speeds, shown in Fig. S13, onto high-resolution topography cannot represent slope-wind effects.) It is questionable whether the free and the forced sublimation should be summed together. In the case of this study, our choice to sum the loss mechanisms to get the total potential sublimation rate does not matter much because the forced sublimation turns out to dominate for the best-fit GCMs. The 2880 × 1440 × 260 grid of $S_{pot}$ values is then annually averaged. The annual-averages are sorted to find the lower and higher values of $S_{pot}$, and each pixel is tagged with a value, $f_{sub}$, between ~0% (most favorable percentage of surface area for snow accumulation) and $f_{sub} \approx$ 100% (least favorable percentage of surface area for snow accumulation). The pixels are cosine-corrected for latitude when doing this interpolation. Cold places are the most favorable for snow accumulation, which occurs below a threshold $f_{snow}$.

Comparing the results of the simplified meltwater model to geologic data.
For each of 100 values of $f_{snow}$ from 1% to 100%, we calculate the predicted melt distribution by finding which pixels satisfy $T_{warmseason} >$ 273 K and also $f_{sub} < f_{snow}$. This gives 5400 forward models to be compared to each of the two time slices of data (early stage rivers and late stage rivers).

The model predictions and the data are spatially masked as described earlier (Fig. 1). Valley networks and fans are binned on an 8 ppd grid. Every pixel containing the center of a valley segment (for >3.6 Ga rivers) or a fan apex (for <3.6 Ga rivers) is assigned as a geologic positive, all other pixels are assigned as a geologic negative. The data are compared to the forward models and the numbers of true positives TP (model+, data+), false positives FP (model+, data-), true negatives TN (model-, data-), and false negatives FN (model-, data-) are calculated. These numbers are cosine-weighted for latitude correction. Metrics for assessing data-model agreement include

- *sensitivity*, the proportion of +ve cases correctly predicted, sensitivity = TP / (TP + FN). (Also known as *recall*.)
- *specificity*, proportion of -ve cases correctly predicted, specificity = TN / (TN + FP).

- *precision,* the fraction of model-predicted positives that are true positives, precision = TP/(TP+FP).

For each of the 54 GCM simulations, we marginalize over the 100 possible values of $f_{snow}$ by plotting two of the above characteristics against each other, varying $f_{snow}$, and then computing a summary statistic. Specifically, for each of the 54 GCM simulations, we calculate a Receiver-Operating Characteristic (ROC) curve varying $f_{snow}$ and measuring the area under the curve (AUC) of a plot of sensitivity against (1-specificity); informedness (Youden's J statistic, ref. *82*); and the precision-recall area-under-curve, PR-AUC). By varying $f_{snow}$ over the full range, we weight all values equally. This is appropriate given ignorance of Early Mars hydrology, but future GCM work with coupled hydrology might motivate truncating at (for example) $f_{snow}$ = 25%. Here we provide details on each summary metric. In practice AUC is calculated using the trapezium rule

$$\text{AUC} = \Sigma_i \left[ (sensitivity_i + sensitivity_{i+1}) / 2 \times (1\text{-}specificity)_{i+1} - (1\text{-}specificity)_i ) \right] \quad (10)$$

$$\text{YoudensJ} = \max(sensitivity - (1\text{-} specificity)) \quad (11)$$

$$\text{PR-AUC} = \Sigma_i \left[ (precision_i + precision_{i+1}) / 2 \times (recall_{i+1} - recall_i ) \right] \quad (12)$$

Our procedure penalizes false positives and false negatives equally. However, an ideal early Mars climate model should have almost no false negatives. This is because (after geologic masking) our river maps are reliable, so false negatives are unlikely to be the result of mapping errors. By contrast, a false positive might be the result of local geologic circumstances de-linking runoff from river incision (e.g., large grain size making sediment transport, and thus landscape modification, difficult). However, penalizing false negatives more harshly than false positives would add an extra adjustable parameter (the punishment ratio) to the model, justifying our omission of this effect.

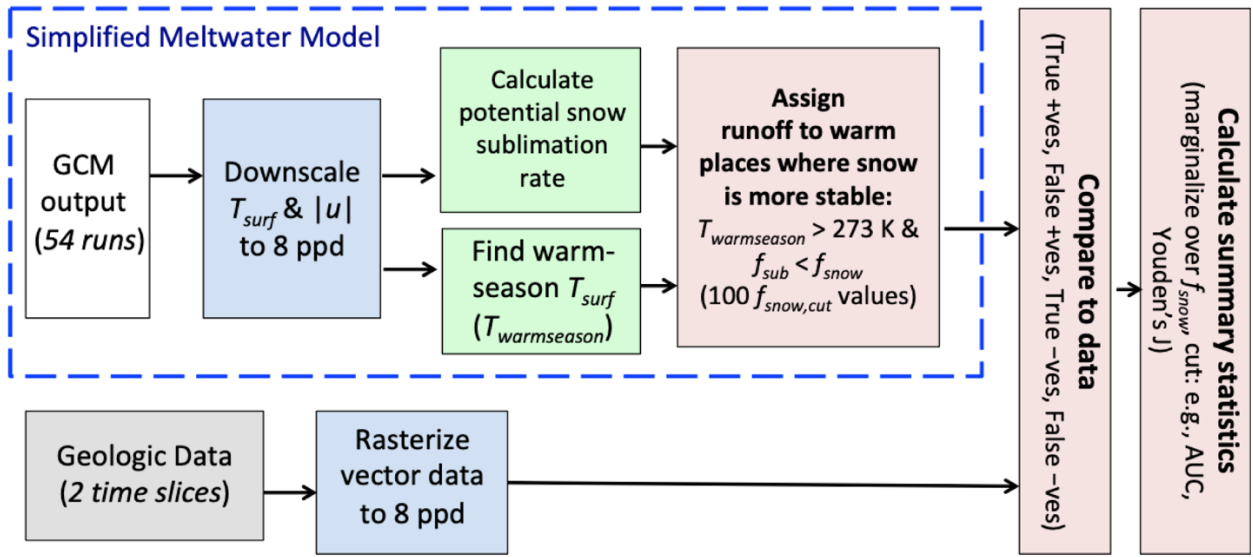

**Fig. S1.** Overview of data-model comparison. Explanation of variables: $T_{surf}$ corresponds to surface temperature, $|u|$ corresponds to surface wind speed, $T_{warmseason}$ corresponds to the $T_{surf}$ of the warmest season, $f_{sub}$ corresponds to the favorability of the pixel for snow accumulation ($f_{sub} = 10\%$ means that only 10% of planet surface area is more favorable), and $f_{snow}$ corresponds to the fraction of planet surface area with snow or ice during the warm season.

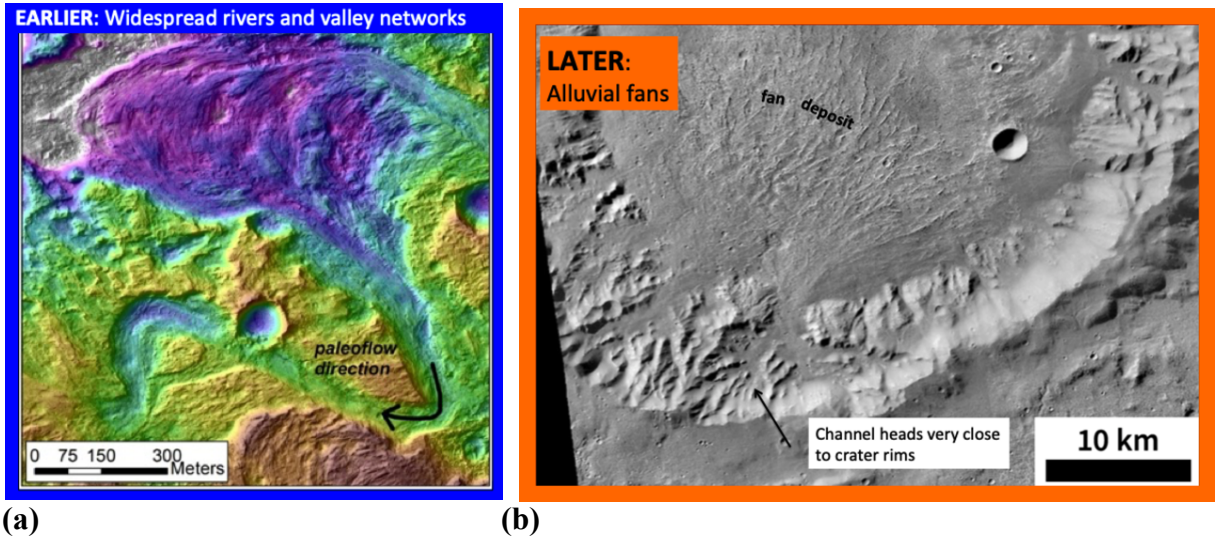

**Fig. S2.** Examples of **(a)** well-preserved fluvial deposits from the Valley Network era (154.64°E 5.37°S), and **(b)** an alluvial fan (location from *11*).

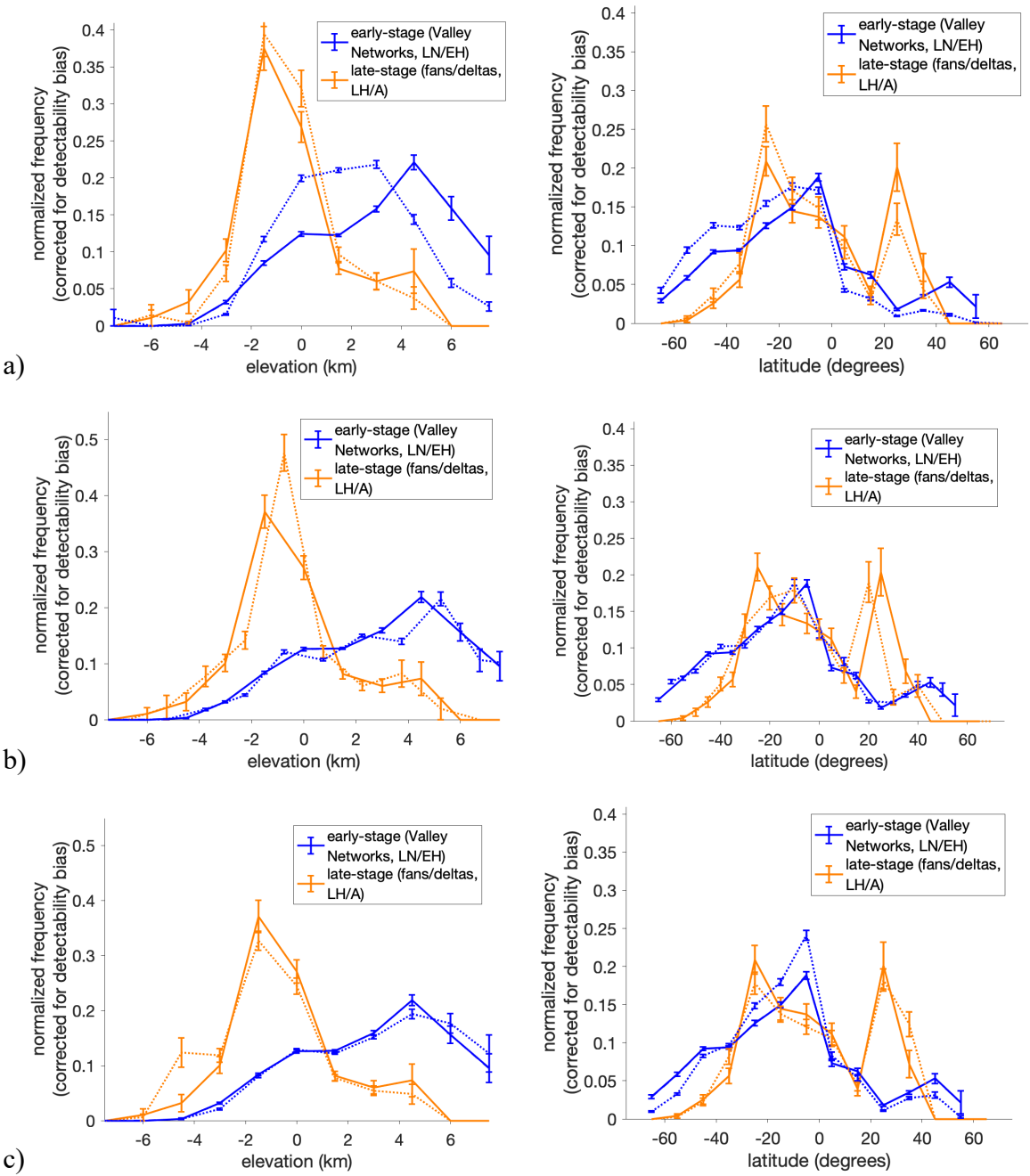

**Fig. S3.** Sensitivity tests for results shown in Fig. 2. Arabia Terra is not excised. **(a)** Uncorrected data shown in dotted lines, debiased data with solid lines. **(b)** Effect of bin offsetting. Dotted lines are with bin offsetting (by ½ bin width), and solid lines are with bins as shown in Fig. 2. **(c)** Solid blue lines: counting Valley Network segment number, instead of weighting by length. Dotted orange lines: 1 km$^2$ cut on the alluvial fans/deltas (instead of 10 km$^2$) and not excising terraced fans. Solid blue/orange lines: as in Fig. 2 (Valley Networks weighted by length, 10 km$^2$ cut on fan/delta size, terraced fans excised).

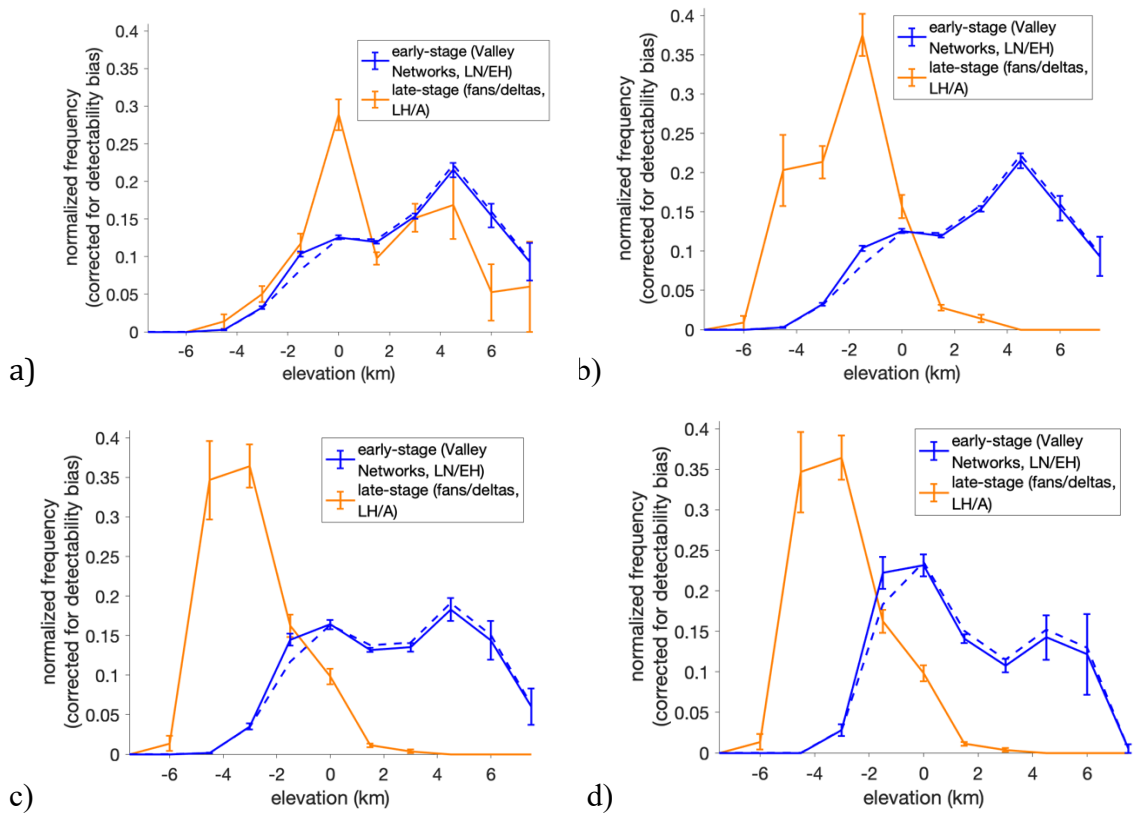

**Fig. S4.** Further sensitivity tests for elevation trends over time. **(a)** Maximum elevation within catchment for fans/deltas, vs. elevation of Valley Network valleys of all Strahler orders (this is not an apples-to-apples comparison, as catchment maximum elevation will overstate the elevation of typical channel heads). **(b)** Fan apex elevation vs. elevation of Valley Network valleys of all Strahler-orders. **(c)** Fan toe elevation vs. elevation of Valley Network valleys of Strahler orders 2 and higher ("trunk" streams). **(d)** Fan toe elevation vs. Strahler orders 3 and higher. Dashed lines correspond to the (small) effect of not excising Arabia Terra.

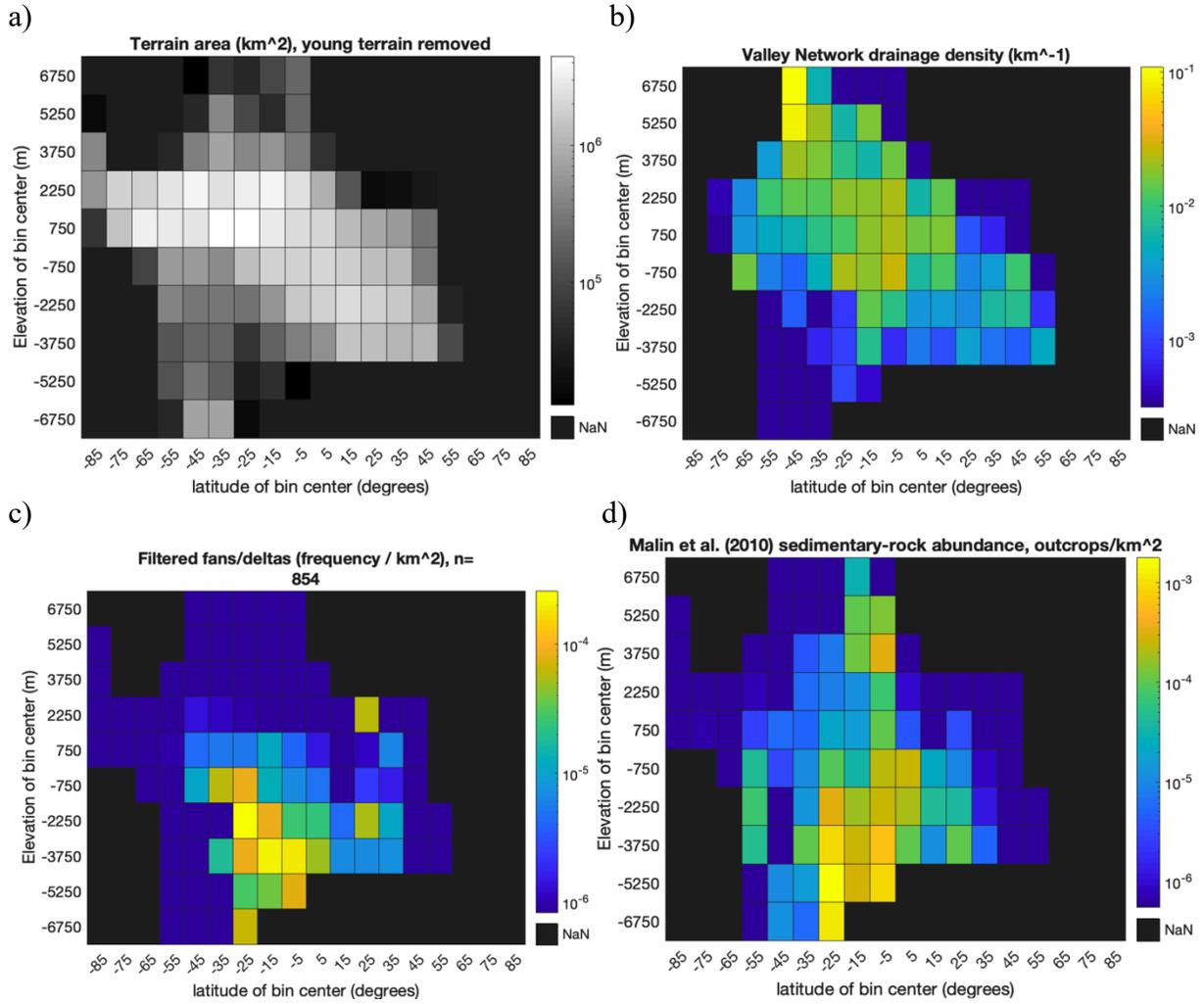

**Fig. S5.** Elevation+latitude distribution of **(a)** non-young terrain (excluding the masked-out area shown in Fig. 1b), **(b)** Valley Networks, **(c)** alluvial fans/deltas, and **(d)** light-toned, layered sedimentary rocks. Late stage latitude dependence is not an artifact of latitude-dependent hyposometry. For this figure only, elevation+latitude pixels with area < $10^4$ km$^2$ are not shown (as they tend to be statistically noisy).

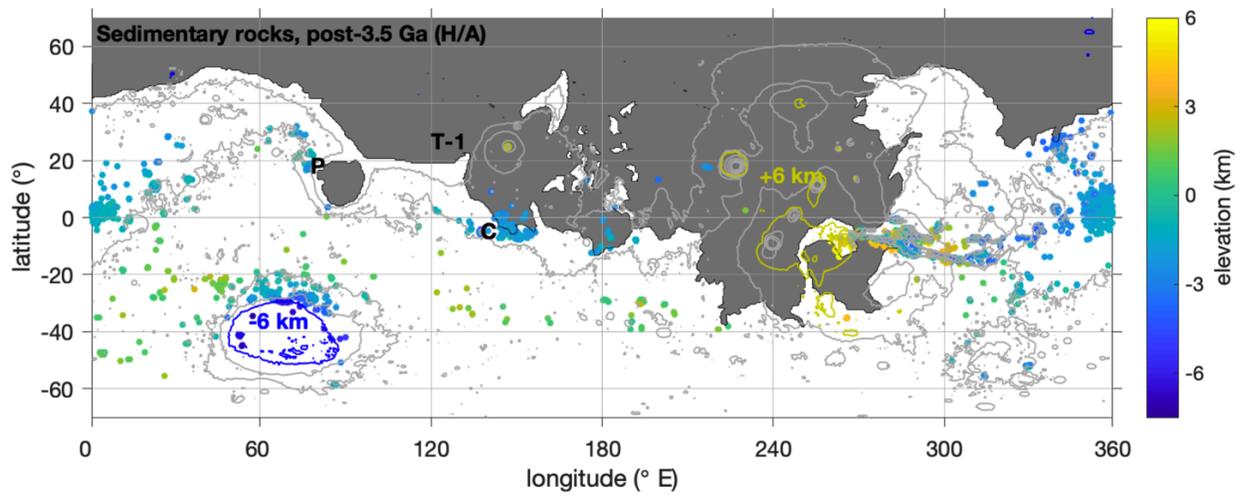

**Fig. S6.** Light-toned layered sedimentary rocks distribution (data from ref. *5*). Contour spacing 3 km. Gray tint: Low/no detection probability. Rovers: C = *Curiosity*, P = *Perseverance*, T-1 = *Tianwen-1* rover (*Zhurong*).

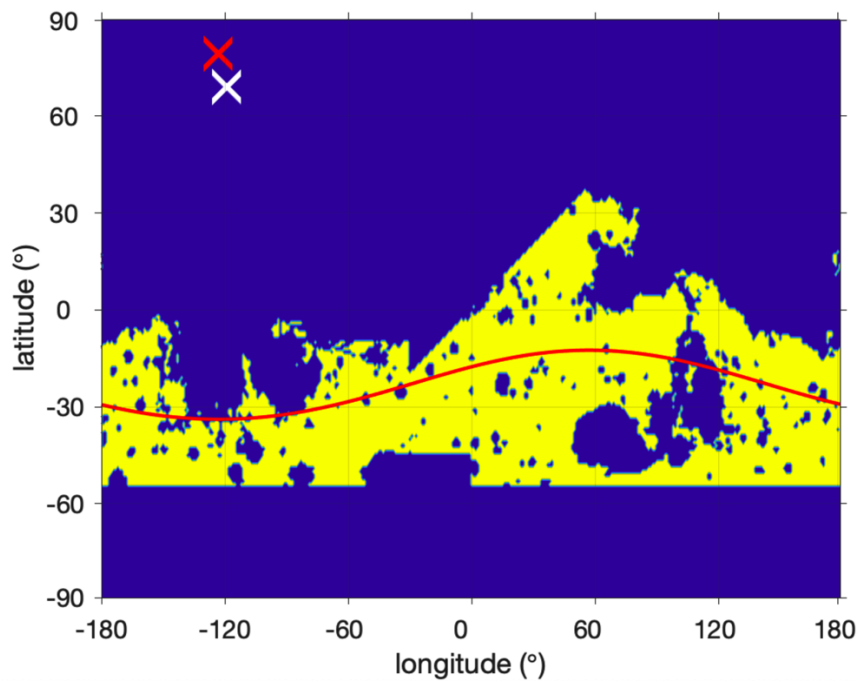

**Fig. S7.** Example of how a small-circle-fitting pipeline can lead to a detection of TPW when no TPW has occurred. We used a synthetic dataset in which, by construction, there is zero evidence for TPW (a uniform distribution). The resulting distribution of Valley Networks is shown in yellow and the resurfacing mask in blue. The red line shows the small circle that is the best fit to the yellow distribution, the corresponding best-fit (spurious) paleopole is shown by the red cross, and the paleopole reported by (*20*) based on Valley Network distribution is shown by the white cross.

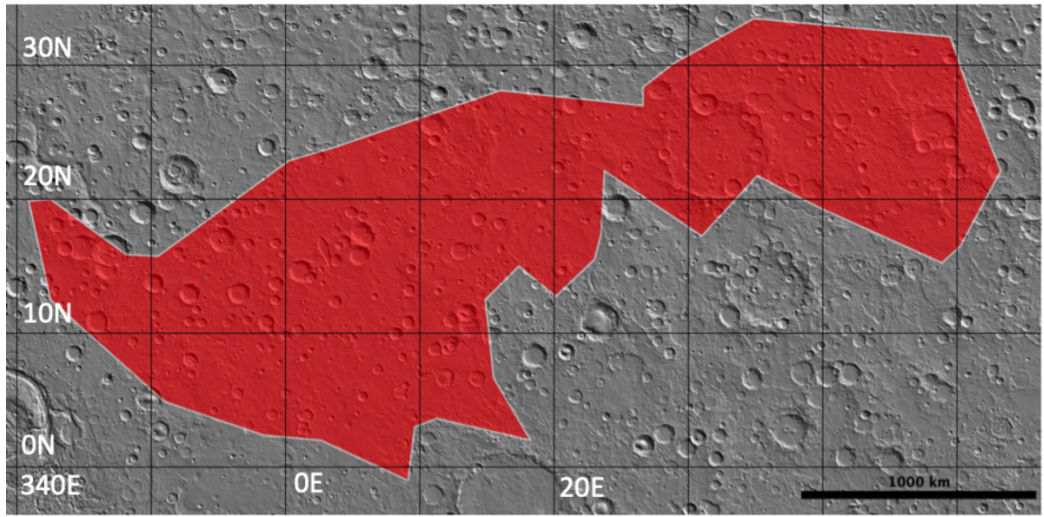

**Fig. S8.** Arabia Terra mask for excluding region of low/no preservation potential for valley networks, guided by the inverted-channel map of ref. *72* (their Fig. S1).

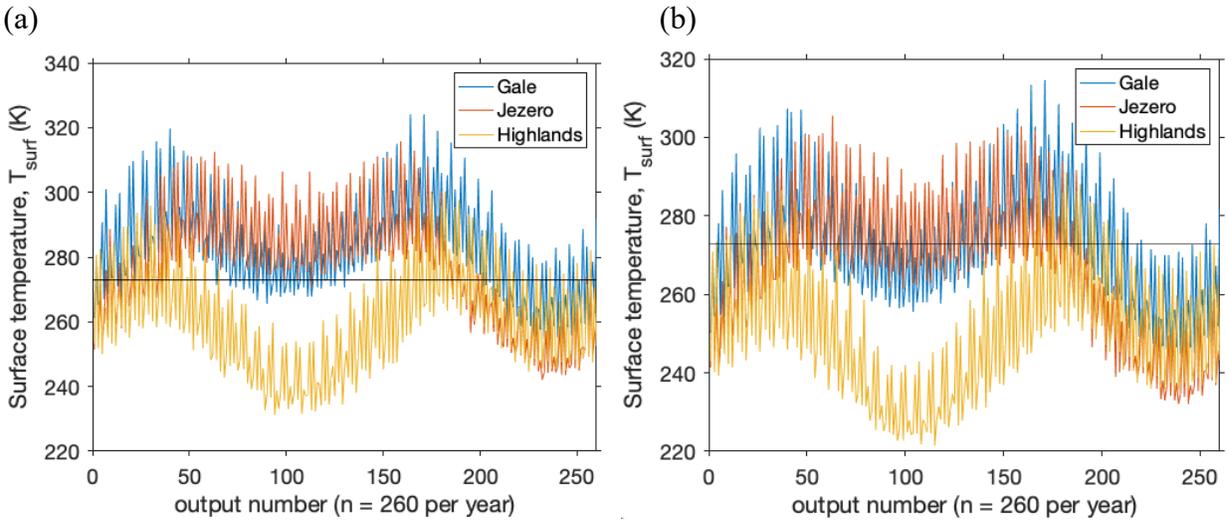

**Fig. S9.** (a) 260-timestep output for GCM run that best matches ~3.6 Ga data ($\tau$ = 3.14, 45° obliquity, $pCO_2$ = 150 mbar). Horizontal line highlights 273 K isotherm. Output is downscaled using MOLA topography. (b) As (a), but for the GCM run that best matches 3.5-3 Ga data ($\tau$ = 2.46, 45° obliquity, $pCO_2$ = 150 mbar). Locations within Gale crater and Jezero crater correspond to rover locations (*Curiosity* and *Perseverance*, respectively).

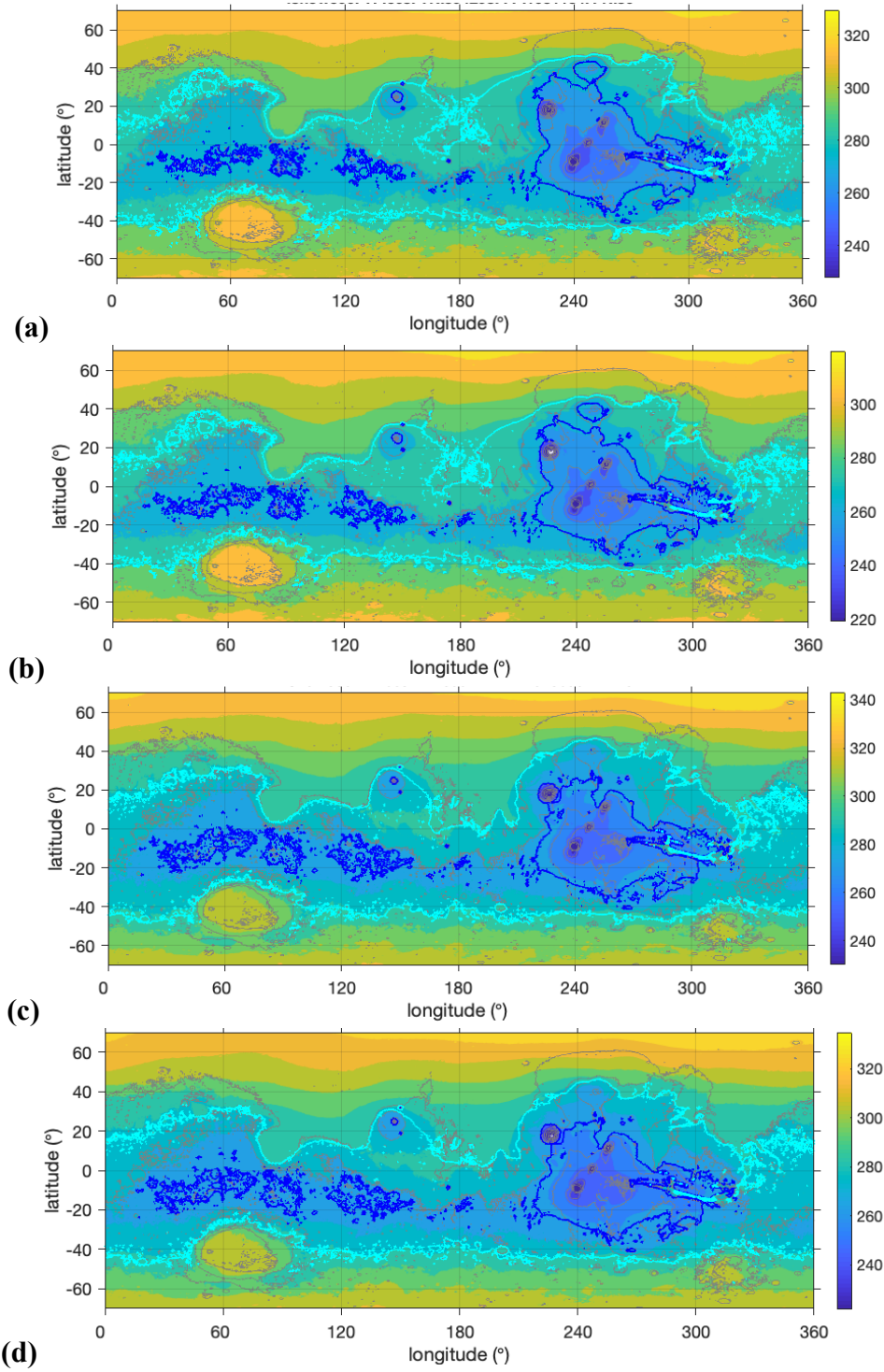

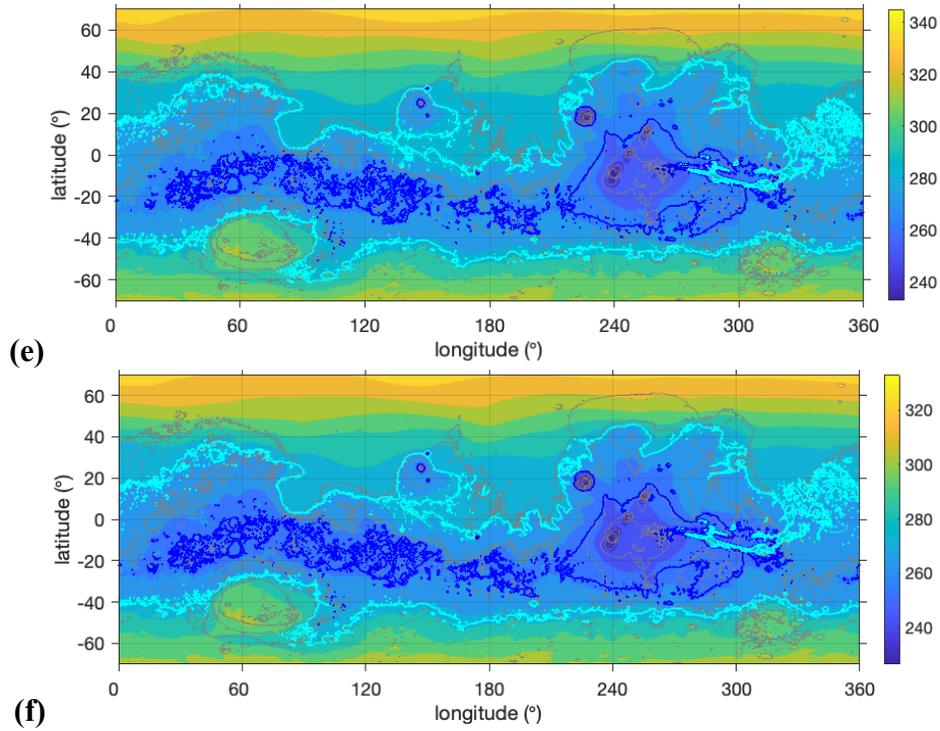

**Fig. S10.** As pressure decreases by a factor of 25 for constant $\tau$, the pattern of warm-season temperatures and snowpack stability does not change much. Figure shows predicted climates for the $\tau$ values corresponding to the overall best fits shown in Figs. 3-5, but varying pCO$_2$ between 500 mbar, 150 mbar and 20 mbar. Colors correspond to warm-season temperature (K) (warmest 100 sols). The dark blue contours contain the part of the planet's surface area with the lowest annually integrated snow sublimation rate (0-10th percentile), and the cyan contours contain the part of the planet's surface area with lower-than-average annually integrated snow sublimation rate (0-50th percentile). Output is downscaled using MOLA topography. Elevation contours (gray) are spaced at 3 km intervals. **(a)** $\tau$ = 3.14, 45° obliquity, pCO$_2$ = 500 mbar. **(b)** $\tau$ = 2.46, 45° obliquity, pCO$_2$ = 500 mbar. **(c)** $\tau$ = 3.14, 45° obliquity, pCO$_2$ = 150 mbar. **(d)** $\tau$ = 2.46, 45° obliquity, pCO$_2$ = 150 mbar. **(e)** $\tau$ = 3.14, 45° obliquity, pCO$_2$ = 20 mbar. **(f)** $\tau$ = 2.46, 45° obliquity, pCO$_2$ = 20 mbar.

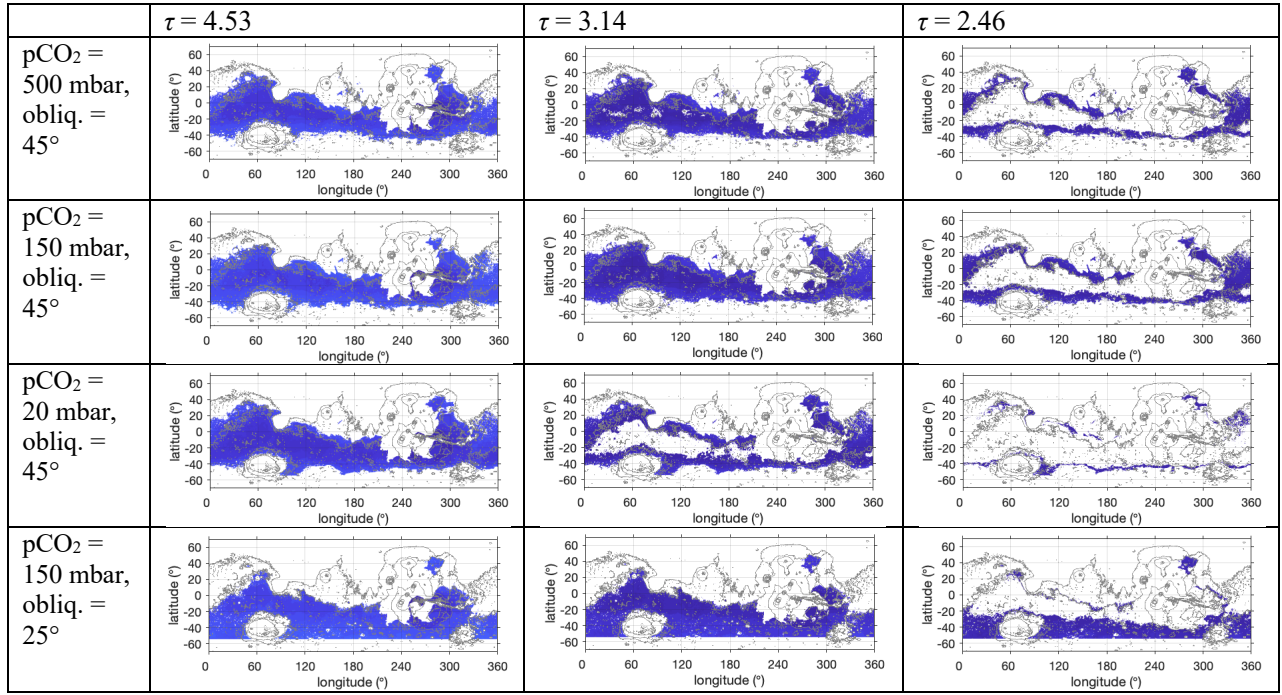

**Fig. S11.** The simplified meltwater predictions are only weakly sensitive to pCO$_2$ (row-wise variation between the first three rows), are strongly sensitive to the strength of non-CO$_2$ greenhouse forcing (column-wise variation), and are strongly sensitive to obliquity (fourth row), but lower obliquity predictions are a poor match to data. Geologically masked output for a fixed $f_{snow}$ of 50%.

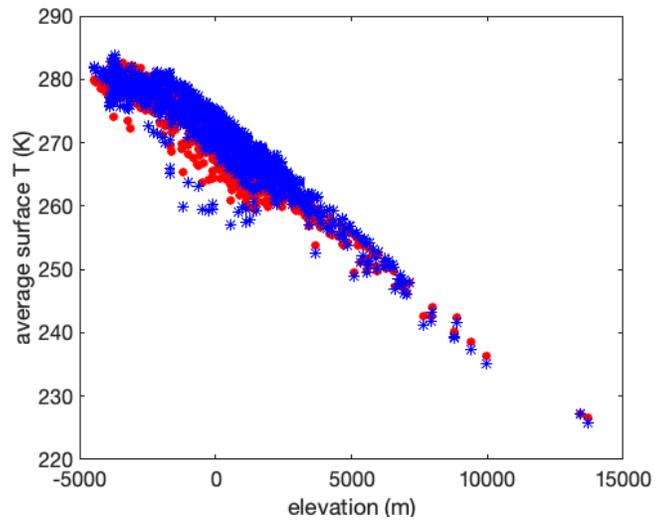

**Fig. S12.** The lapse rate in average surface temperature for a 500 mbar run (run 908, red dots) is almost the same as the lapse rate in average surface temperature for a 200 mbar run with the same non-$CO_2$ greenhouse forcing and obliquity (run 218, blue asterisks).

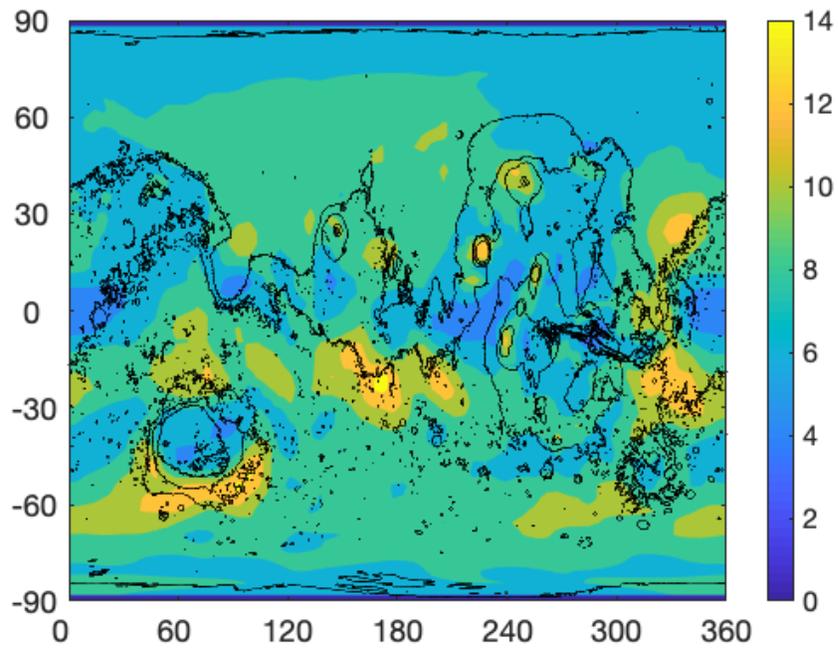

**Fig. S13.** Annual-average wind speed at the lowest level in the atmosphere, in m/s (for the run shown in Fig. 3a). Black contours correspond to topography and are drawn at 3 km intervals starting from -6 km.

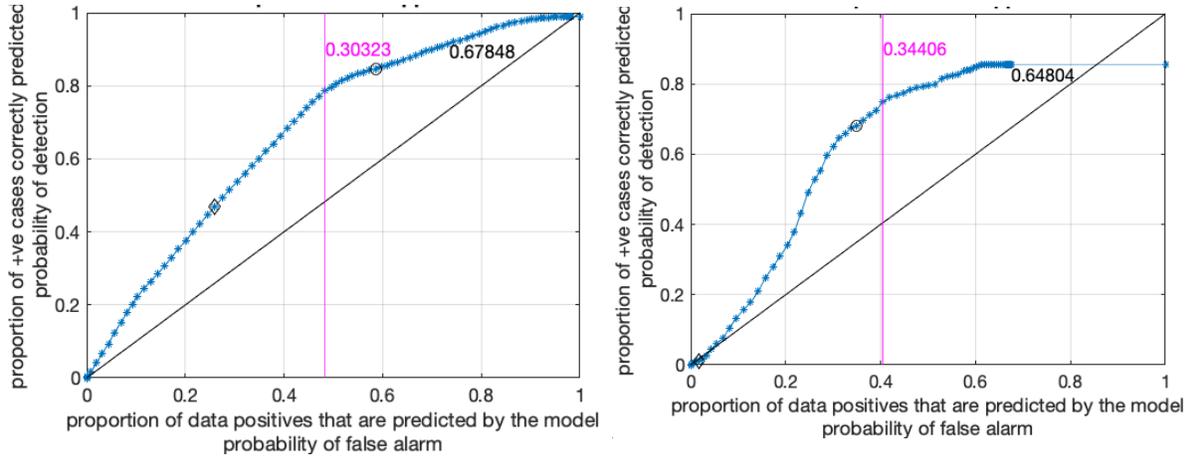

**Fig. S14.** Receiver operating characteristic (ROC) curves for (left) comparison of the best-fit model for >3.6 Ga data ($\tau = 3.14$, 45° obliquity, $pCO_2 = 150$ mbar) to those early-stage data, (right) comparison of the best-fit model for <3.6 Ga data $\tau = 2.46$, 45° obliquity, $pCO_2 = 150$ mbar to those late-stage data. Points plotting above the diagonal black line correspond to models that perform better than chance. The blue asterisks are drawn at intervals of 1% in $f_{snow}$. The black diamond highlights the run at $f_{snow} = 25\%$ (in each panel) and the black circle highlights the run at $f_{snow} = 50\%$ (in each panel). The vertical purple line marks the best-fit $f_{snow}$ (with the corresponding Youden's J value in magenta). The black value in the top right is the ROC Area Under Curve (AUC).

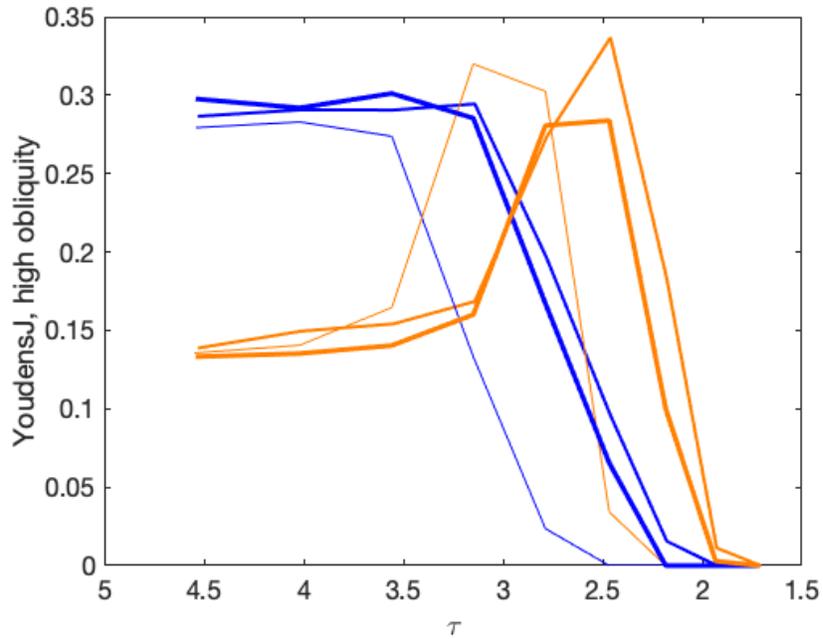

**Fig. S15.** Youden's J as a function of $\tau$, comparing models to alluvial fan data (orange lines) and Valley Network data (blue lines). The thickest lines are for $pCO_2$ = 500 mbar, the thinnest lines are for $pCO_2$ = 20 mbar, and the intermediate lines are for $pCO_2$ = 150 mbar. The data-model comparison is relatively insensitive to $pCO_2$, and strongly sensitive to $\tau$. Figure shows results for 45° obliquity.

| Run ID | Input parameters | | | Output | Summary data-model comparison scores | | | |
|---|---|---|---|---|---|---|---|---|
| | $pCO_2$ (mbar) | $\tau$ | Obliq. (°) | (K) | AUC, VN era | AUC, fans era | YoudensJ, VN era | YoudensJ, fans era |
| 215 | 150 | 4.53 | 45 | 284.8 | 0.6669 | 0.4936 | 0.2866 | 0.1387 |
| 216 | 150 | 4.01 | 45 | 278.9 | 0.6678 | 0.5031 | 0.2907 | 0.1496 |
| 217 | 150 | 3.55 | 45 | 273.3 | 0.6702 | 0.5034 | 0.2906 | 0.1542 |
| 218 | 150 | 3.14 | 45 | 267.8 | 0.6725 | 0.5211 | 0.2944 | 0.1685 |
| 219 | 150 | 2.78 | 45 | 262.7 | 0.5579 | 0.6165 | 0.1960 | 0.2735 |
| 220 | 150 | 2.46 | 45 | 257.9 | 0.3898 | 0.6613 | 0.0959 | 0.3367 |
| 225 | 20 | 4.54 | 45 | 272.7 | 0.6635 | 0.4645 | 0.2793 | 0.1356 |
| 226 | 20 | 4.02 | 45 | 267.1 | 0.6660 | 0.4711 | 0.2829 | 0.1406 |
| 227 | 20 | 3.56 | 45 | 261.7 | 0.6570 | 0.4967 | 0.2738 | 0.1646 |
| 228 | 20 | 3.15 | 45 | 256.5 | 0.4796 | 0.6413 | 0.1333 | 0.3199 |
| 229 | 20 | 2.79 | 45 | 251.5 | 0.2681 | 0.5905 | 0.0236 | 0.3025 |
| 230 | 20 | 2.47 | 45 | 248.2 | 0.1331 | 0.2320 | 0 | 0.0342 |
| 315 | 150 | 4.53 | 25 | 289.0 | 0.5656 | 0.3656 | 0.1022 | 0.0433 |
| 316 | 150 | 4.01 | 25 | 283.0 | 0.5656 | 0.3652 | 0.1014 | 0.0542 |
| 317 | 150 | 3.55 | 25 | 277.2 | 0.5655 | 0.3717 | 0.0977 | 0.0575 |
| 318 | 150 | 3.14 | 25 | 271.6 | 0.5641 | 0.3749 | 0.0930 | 0.0655 |
| 319 | 150 | 2.78 | 25 | 266.3 | 0.4880 | 0.4348 | 0.0201 | 0.1252 |
| 320 | 150 | 2.46 | 25 | 261.1 | 0.3485 | 0.5554 | 0 | 0.2529 |
| 325 | 20 | 4.54 | 25 | 277.5 | 0.4766 | 0.3416 | 0.0682 | 0.0064 |
| 326 | 20 | 4.02 | 25 | 271.4 | 0.4756 | 0.3436 | 0.0692 | 0.0127 |
| 327 | 20 | 3.56 | 25 | 265.5 | 0.4668 | 0.3640 | 0.0600 | 0.0326 |
| 328 | 20 | 3.15 | 25 | 259.8 | 0.3733 | 0.4935 | -0.0003 | 0.1665 |
| 329 | 20 | 2.79 | 25 | 254.1 | 0.2038 | 0.5411 | -0.0003 | 0.2352 |
| 330 | 20 | 2.47 | 25 | 248.3 | 0.0493 | 0.2178 | -0.0003 | -0.0000 |
| 611 | 150 | 2.18 | 45 | 253.3 | 0.2075 | 0.4577 | 0.0156 | 0.1830 |
| 612 | 150 | 1.93 | 45 | 248.9 | 0.1329 | 0.1776 | 0 | 0.0113 |
| 613 | 150 | 1.71 | 45 | 245.0 | 0.1103 | 0.0780 | 0 | 0 |
| 621 | 20 | 2.18 | 45 | 241.2 | 0.1030 | 0.0579 | 0 | 0 |
| 622 | 20 | 1.93 | 45 | 236.4 | 0.0755 | 0.0160 | 0 | 0 |
| 623 | 20 | 1.71 | 45 | 232.1 | 0.0545 | 0.0028 | 0 | 0 |
| 711 | 150 | 2.18 | 25 | 256.4 | 0.1983 | 0.5167 | 0 | 0.2222 |
| 712 | 150 | 1.93 | 25 | 251.8 | 0.0610 | 0.2074 | 0 | 0.0110 |
| 713 | 150 | 1.71 | 25 | 247.4 | 0.0284 | 0.0639 | 0 | 0 |
| 721 | 20 | 2.18 | 25 | 242.6 [†] | 0.0173 | 0.0418 | 0 | 0 |
| 722 | 20 | 1.93 | 25 | 236.9 [†] | 0.0004 | 0.0012 | 0 | 0 |
| 723 | 20 | 1.71 | 25 | 231.7 [†] | 0 | 0 | 0 | 0 |
| 905 | 500 | 4.54 | 45 | 286.3 | 0.6694 | 0.4663 | 0.2976 | 0.1332 |
| 906 | 500 | 4.02 | 45 | | 0.6723 | 0.4744 | 0.2919 | 0.1353 |
| 907 | 500 | 3.56 | 45 | 275.3 | 0.6772 | 0.4906 | 0.3011 | 0.1404 |
| 908 | 500 | 3.15 | 45 | 270.2 | 0.6623 | 0.5108 | 0.2854 | 0.1601 |
| 909 | 500 | 2.79 | 45 | 265.5 | 0.5209 | 0.6331 | 0.1672 | 0.2806 |
| 910 | 500 | 2.47 | 45 | 260.8 | 0.3533 | 0.6104 | 0.0655 | 0.2839 |
| 951 | 500 | 2.18 | 45 | 256.4 | 0.1899 | 0.3580 | 0.0000 | 0.1000 |
| 952 | 500 | 1.93 | 45 | 252.3 | 0.1388 | 0.1555 | 0 | 0.0029 |
| 953 | 500 | 1.71 | 45 | 248.3 [**] | 0.1054 | 0.0677 | 0 | 0 |
| 1005 | 500 | 4.54 | 25 | 290.3 | 0.5449 | 0.3363 | 0.0882 | 0.0350 |
| 1006 | 500 | 4.02 | 25 | 284.8 | 0.5497 | 0.3399 | 0.0903 | 0.0430 |
| 1007 | 500 | 3.56 | 25 | 279.1 | 0.5509 | 0.3460 | 0.0922 | 0.0487 |
| 1008 | 500 | 3.15 | 25 | 273.8 | 0.5534 | 0.3540 | 0.0915 | 0.0594 |
| 1009 | 500 | 2.79 | 25 | 268.8 | 0.4495 | 0.4488 | 0.0016 | 0.1594 |
| 1010 | 500 | 2.47 | 25 | 264.0 | 0.3216 | 0.5264 | 0 | 0.2358 |
| 1051 | 500 | 2.18 | 25 | 259.3 | 0.1634 | 0.4425 | 0 | 0.1744 |
| 1052 | 500 | 1.93 | 25 | 255.0 | 0.0473 | 0.1862 | 0 | 0.0325 |
| 1053 | 500 | 1.71 | 25 | 250.9 | 0.0183 | 0.0395 | 0 | 0 |
| 401 | 1000 | 0 | 45 | 209.1 | - | - | - | - |
| 411 | 150 | 0 | 45 | 203.6 | - | - | - | - |
| 421 | 20 | 0 | 45 | <200 | - | - | - | - |

**Table S1.** Summary of GCM output and data-model comparison.
[**] Possibly still cooling at the end of run (and much too cold to match data).
[†] Some of these extremely cold 20 mbar runs show evidence for ongoing atmospheric collapse.